\documentclass[noshowpacs,prl,twocolumn,showpacs,preprintnumbers,amsmath,aps,superscriptaddress]{revtex4-1}

\usepackage{graphicx,hyperref}% Include figure files
\usepackage{bm}% bold math
\usepackage{subfigure}% bold math

\begin{document}

\title{Perspective: The Glass Transition}

\author{Giulio Biroli} \email{giulio.biroli@cea.fr}
\affiliation{IPhT, CEA/DSM-CNRS/URA 2306, CEA Saclay, F-91191 Gif-sur-Yvette Cedex, France}

\author{Juan P. Garrahan} \email{juan.garrahan@nottingham.ac.uk}
\affiliation{School of Physics and Astronomy, University of Nottingham, Nottingham, NG7 2RD, United Kingdom}

\date{\today}

\begin{abstract}
We provide here a brief perspective on the glass transition field.  It is an assessment, written from the point of view of theory, of where the field is and where it seems to be heading.  We first give an overview of the main phenomenological characteristics, or ``stylised facts,'' of the glass transition problem, i.e.\ the central observations that a theory of the physics of glass formation should aim to explain in a unified manner.  We describe recent developments, with a particular focus on real space properties, including dynamical heterogeneity and facilitation, the search for underlying spatial or structural correlations, and the relation between the thermal glass transition and athermal jamming.  We then discuss briefly how competing theories of the glass transition have adapted and evolved to account for such real space issues.  We consider in detail two conceptual and methodological approaches put forward recently, that aim to access the fundamental critical phenomenon underlying the glass transition, be it thermodynamic or dynamic in origin, by means of biasing of ensembles, of configurations in the thermodynamic case, or of trajectories in the dynamic case.  We end with a short outlook.
\end{abstract}

\maketitle

\section{Introduction}

The aim of this paper is to provide a brief assessment of the current state of the glass transition field, specifically to highlight what in our view are the more promising recent developments and what progress we expect (or hope) to see in the near future in this area of research.  We do not aim to give in any way a comprehensive coverage---there are many extensive reviews such as those in Refs.\ \cite{Angell1995,Ediger1996,Angell2000,Debenedetti2001}
and the more recent ones in Refs.\ \cite{Lubchenko2007,Heuer2008,Cavagna2009,Chandler2010,Berthier2011}---but rather a perspective on what we perceive to be the central questions in the field and the avenues that we think should be pursued to answer them.  This perspective is mostly theoretical and it is heavily biased by our own works and overall approaches to the glass transition problem.  

In its most general sense the glass transition refers to the generic change in a many-body system from an equilibrium fluid state to a non-equilibrium disordered solid state \cite{Angell1995,Ediger1996,Angell2000,Debenedetti2001,Lubchenko2007,Heuer2008,Cavagna2009,Chandler2010,Berthier2011}.  This change is not a transition in the thermodynamic sense, at least in what is observed in experiments, but a kinetic phenomenon where the amorphous solid is dynamically arrested, i.e.\ does not have enough time to relax on experimental timescales.  The basic physical ingredients of the glass transition are those of a many-body system with excluded volume, or similarly frustrating, local interactions.  The change from the fluid to the amorphous solid is brought about by an effective increase in density which renders relaxation to the true stable thermodynamic state, typically a crystal, impractical on the available observation time, leaving the system trapped in a disordered yet solid metastable state.  The prototypical glass formers are molecular liquids supercooled beyond their crystallisation transitions, where the decrease in temperature leads to an effective increase in density. Glass transitions are also observed over a range of scales in particulate systems for essentially the same reasons, ranging from colloidal suspensions and granular materials to cellular cultures.  Glass is a generic state of matter and glass transitions a common occurrence in many-body systems.  This is why understanding this problem is so important in condensed-matter science. 

If we compare the state of the glass transition field as it is now \cite{Cavagna2009,Chandler2010,Berthier2011} with how it was, say, a decade ago 
\cite{Angell2000,Debenedetti2001}, it is evident there has been a substantial shift of focus away from what we can broadly call ``landscape'' concepts and methods towards real space (or real space and time) properties.  Ten years or so ago, a dominant strand of thinking followed the idea that glasses could be understood in terms of the properties of the complex energy surface in configuration space on which these systems had to evolve; studies on the statistical properties of local minima, or inherent structures, abounded, and theory focussed on understanding the relations between these global properties and the observed phenomenology, as summarised in this highly regarded review from that time \cite{Debenedetti2001}.  The shift towards real space, as opposed to configuration space, properties, occurred in great part due to the experimental (and numerical) discovery of dynamic heterogeneity \cite{Ediger2000,Glotzer2000,Andersen2005}, which forced theory in turn to consider seriously the role of fluctuations in (real) space and time.  This shift has occurred both in the evolution of landscape-based approaches \cite{Lubchenko2007,Berthier2011} and in the emergence of new perspectives \cite{Chandler2010}.  In our discussion below we will focus mainly on this new emphasis on real space concepts and methods.

The paper is organised as follows.  In the next section we review what in our view are the central phenomenological observations that characterise the glass transition.  Borrowing a terminology from economics, we call them the ``stylised facts'' of the glass problem.  The development of a successful theory should strive to explain them in as unified a manner as it is possible.  We then proceed to the main sections of the paper, where we discuss recent developments including dynamic heterogeneity and its connection of dynamic facilitation, the search for static correlations, and the relation between the glass transition and jamming.  We also discuss similarities and differences between the main conceptual approaches to the glass problem, and how the search for clear evidence that validates the different competing views is driving the development of new methodologies which may end up having wider applicability.  We finish with an outlook of where we see the field going and of new avenues of research that seem to be opening up.

\begin{figure*}[th]
\includegraphics[width=2\columnwidth]{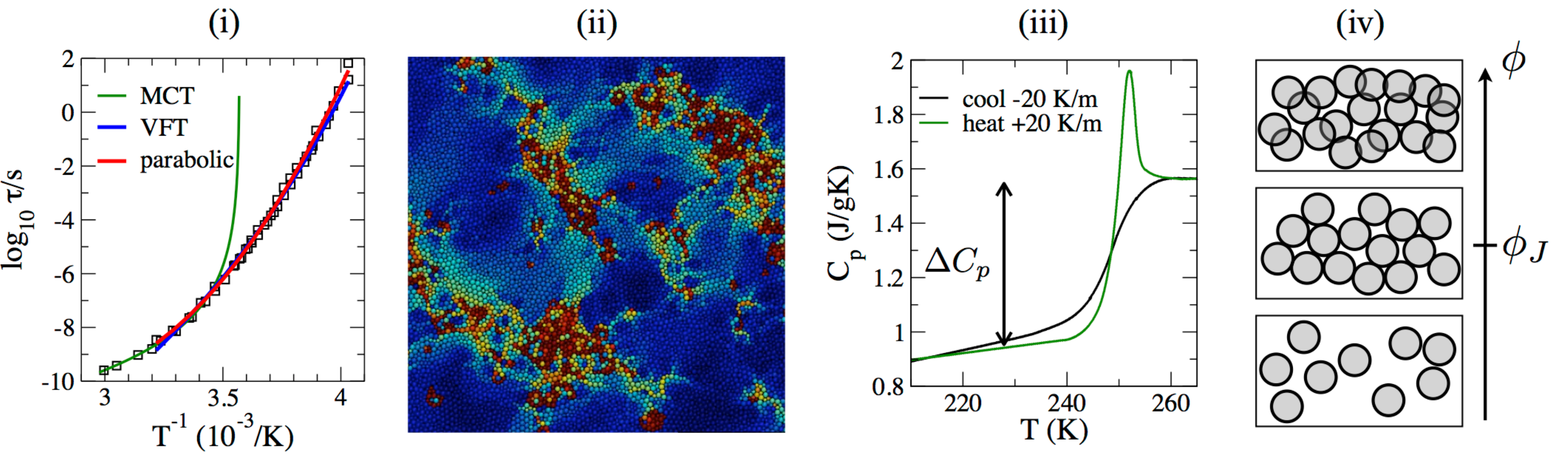}
\caption{Stylised facts of the glass transition. (i) Slowdown with no apparent structural change: structural relaxation time $\tau$ for OTP (symbols, data from \cite{Richert1998,Gottke2001,Richert2005}); the blue curve is a VFT fit, $\log{\tau / \tau_0} = A / (T-T_c)$ \cite{Richert2005}; the red curve is a fit with the parabolic law, $\log{\tau / \tau_0} = (J/T_o)^2 (T_o/T -1)^2$; the red curve is an MCT fit, $\tau \propto |T-T_{\rm mct}|^{-\gamma}$.  (See \cite{Richert2005,Elmatad2009,Gottke2001} for details on these fits.) (ii) Dynamical heterogeneity: projection in space of an equilibrium trajectory of a two-dimensional supercooled mixture, from Ref.\ \cite{Keys2011}; particles coloured according to overlap with initial positions (displacement by a particle diameter or more is dark red, and no displacement is dark blue); the trajectory length is about a tenth of a relaxation time at these conditions; spatial segregation of dynamics is evident. (iii) Anomalous thermodynamic response: temperature variation of the specific heat $C_p$ of OTP on cooling (black curve) and heating (green curve), from differential scanning calorimetry \cite{Velikov2001}; $\Delta C_p$ is the difference in specific heat between the liquid and the glass.  (iv) Jamming at zero temperature: for densities below $\phi_J$ particles are not in contact and the system is fluid; at the jamming density $\phi_J$ the system becomes isostatic and mechanically stable; for densities beyond $\phi_J$ particles would overlap, a situation not allowed for hard objects.
\label{Fig1}}
\end{figure*}

\section{Stylised facts of the glass transition}

Loosely speaking, in conventional condensed-matter systems structure determines dynamics: e.g.\ liquids are disordered and thus flow and relax (in the sense that they decorrelate from their initial conditions), while crystals are ordered and do not (although whether crystals ``flow'' is a moot point \cite{Sausset2010}).  In particular, sudden changes in dynamical behaviour follow from similar sudden changes in structure, such as those due to thermodynamic phase transitions as in the example of liquid to crystal.  But the glass transition does not fit within this paradigm in an obvious way.  
A supercooled liquid slows down, to the point of complete arrest, while at the same time maintaining its liquid structure.  This leads to what is probably the fundamental question in the field: is the glass transition as observed experimentally purely a dynamical phenomenon where the fluid becomes kinetically arrested, or is the observed dynamics the consequence of an underlying phase transition from the fluid to a thermodynamic glass state?  From the point of view of theory, the central schism is given by how this question is answered.  For example, in our writing on the glass problem one of us has mostly advocated and developed a fundamentally thermodynamic perspective, the so-called random first-order transition theory (RFOT) \cite{KTW}, while the other a fundamentally kinetic one, dynamic facilitation (DF) theory; we elaborate on these two perspectives, and others, below.  At this moment in time, neither experiments nor simulations are able to provide a conclusive answer as to whether the glass transition is at its core a thermodynamic or a dynamic phenomenon, so from the theory angle at least there is everything still to play for.   

The glass transition is accompanied by a number of phenomenological characteristics or observations which any satisfactory theory should aim to explain in a unified way.  Of these ``stylised facts'' of glass formation the four central ones, in our view, are the following: (i) slowdown without apparent structural change, (ii) dynamical heterogeneity, (iii) anomalous thermodynamic response, and (iv) mechanical stability at jamming.  The first two relate to how dynamics becomes increasingly cooperative and complex on approach to dynamic arrest from the equilibrium side, for example as a supercooled liquid is cooled towards the glass.  Both slowness and heterogeneity remain prevalent also in the glass state, for example during the slow non-equilibrium drift called aging, but since they initially manifest in equilibrium dynamics it is in this regime that they should be explained first.  Fact (iii) is the salient characteristic of glass formers as they are driven out-of-equilibrium across the experimental glass transition.  It refers to how thermodynamic quantities respond to the system being driven into or out of the glass state.  It provides the key observations on the interplay between thermodynamics and dynamics, and is therefore central to the question of whether the glass transition is essentially thermodynamic or not.  One can argue that fact (iv) is the most out-of-equilibrium of all four. It is about systems with excluded volume interactions at zero temperature, at or close to the density where, while still disordered, they become load bearing and develop a yield stress.  This relates to the connection between the athermal transition to mechanical stability or ``jamming'' and the thermal glass transition.  

Figure 1 illustrates the stylised facts.  Fig.\ 1(i) shows as an example the relaxation time of ortho-therphenyl \cite{Richert1998,Gottke2001,Richert2005}, or OTP, an organic liquid which has been widely studied experimentally in its supercooled regime due to its ease for glass formation.  Over a small range of temperatures the relaxation time grows by many orders of magnitude, eventually reaching $100s$ which conventionally defines the glass transition for liquids (as the corresponding viscosity is so large it is not realistic to distinguish such a sluggish liquid from a solid).  This growth of primary relaxation time or viscosity is characteristic of glass forming liquids, and is what eventually forces the system out of equilibrium on experimental timescales.  A central feature is that the timescale grows with decreasing temperature in a super-Arrhenuis, i.e.\ faster than an exponential of inverse temperature, manner, indicating that the slowdown is the consequence of collective effects.  One of the central aims of theory over the years has been to uncover the functional form of relaxation laws.  We show in the figure fits to the data of three such forms.  One (blue curve) is the Vogel-Fulcher-Tammann (VFT) law, $\log{\tau/\tau_0} = A/(T - T_c)$ \cite{Angell2000}, and a second one (red curve) is the parabolic law $\log{\tau / \tau_0} = (J/T_o)^2 (T_o/T -1)^2$ \cite{Elmatad2009}.  
The two functions fit the data within the shown range but they have very different physical basis.  The VFT form assumes a singularity at some non-zero $T_c$, whose origin can be justified within the RFOT in terms of an ideal thermodynamic transition \cite{Lubchenko2007,Cavagna2009,Berthier2011}; the parabolic form is only singular at $T=0$, and is the relaxation form predicted for hierarchical dynamics within DF theory \cite{Chandler2010}.  That both, despite their completely distinct theoretical underpinning, account reasonably well for the relaxation time of most measured liquids in the deeply supercooled regime highights the fact that a single or a small number of experimental observations are not enough---yet---to distinguish between leading theories.  In the figure we also show a fit \cite{Gottke2001} with a power law form, as predicted by mode-coupling theory (MCT) \cite{Gotze1992,Reichman2005}, $\tau \propto |T-T_{\rm mct}|^{-\gamma}$.  MCT captures the initial stages of supercooling but predicts a divergence at $T_{\rm mct}$ which is not observed.  In this case the phenomenology clearly points towards a change in physical mechanism with decreasing temperature that is not captured by MCT, but which theories such as RFOT and DF should be able to explain.  (As we discuss below, MCT can be incorporated into RFOT, while for DF the failure of MCT is an indication of the onset of true glassiness.)

Figure 1(ii) exemplifies dynamic heterogeneity.  It shows the projection in space of the equilibrium dynamics of a two-dimensional supercooled mixture (from Ref.\ \cite{Keys2011}).  Particles are coloured according to their overlap with their initial positions: a particle that is displaced by more than one particle diameter is dark red; a particle that has no displacement is dark blue; intermediate colours coincide with intermediate displacements.  Highly mobile particles are clustered in space, as are highly immobile ones.  The figure illustrates the fact that relaxation is heterogeneous, both in time and space, and that spatial correlations build up in the dynamics over length scales much larger than those apparent from structure, which to all intents and purposes is the same as that of the normal liquid.  As a system gets progressively slow, with decreasing temperature or increasing density, dynamic heterogeneity gets more pronounced.  It has now been observed experimentally in virtually all kinds of systems that undergo glass transitions, including molecular liquids \cite{Ediger2000}, colloids \cite{Weeks2000,Kegel2000}, granulars \cite{Dauchot2005,Keys2007}, aging systems \cite{Mayer2004,Brun2013}, and dense living matter \cite{Angelini2011}.  
An important consequence of dynamic heterogeneity is that the relations between transport coefficients that hold in the normal liquid state, typically derived under assumptions of homogeneity, break down in the supercooled regime \cite{Schweizer2007}.  

The anomalous behaviour of thermodynamic response functions is illustrated in Fig.\ 1(iii).  It shows, again for the example of OTP, the characteristic hysteresis in the specific heat upon cooling/heating the supercooled liquid into/from the non-equilibrium glass as measured by differential scanning calorimetry  \cite{Velikov2001}.  The figure shows cooling/heating at the ``standard'' rate of $20 K/{\rm min}$.  Upon cooling the specific heat drops from a liquid-like value to a solid-like, as one would expect by the loss of fluctuations as the liquid becomes solid.  This  drop, $\Delta C_p$, is sometimes considered an important signature of the thermodynamic basis for glass formation \cite{Angell2000,Lubchenko2007,Berthier2011}, although it also possible to interpret it purely in terms of elastic responses in the liquid \cite{Trachenko2011,wyartcp}.  Another significant feature is the asymmetry between cooling and heating: upon heating the specific heat displays a peak before restoring to the liquid value.  At the standard heating rate, the temperature at which the upturn of the heat capacity occurs is used as an alternative experimental definition of the glass transition, as this temperature often coincides with that defined via the relaxation time reaching 100s.  The asymmetry in the specific heat is a clear indication that the liquid to glass transition observed under experimental conditions is an out-of-equilibrium phenomenon.  Other thermodynamic responses, such as specific volume or refractive index, show analogous behaviour.  An important question is whether these observations are a non-equilibrium precursor to a true thermodynamic transition between the liquid and an ideal glass state, for example occurring in the ideal quasistatic limit, or can be understood on purely dynamical grounds by how the liquid responds to becoming arrested.

Figure 1(iv) illustrates the fourth of our stylised facts, the onset of mechanical stability in a disordered collection of hard (or semi-hard) objects, or ``jamming'' \cite{Van-Hecke2009,Liu2010,procaccia,Torquato2010,barrat}.  Consider a system of hard particles, such as the disks of Fig.\ 1(iv).  At low density particles typically are not in contact as there is enough free volume between them, a situation that one would naturally associate with a fluid state.  On increasing density the free volume will be reduced, and at some point the most likely situation is that a particle is in contact with its neighbours.  Eventually there are just enough contacts to allow for mechanical stability, and the system becomes isostatic (the precise number of contacts per particle depends on the specific shape of the objects and the details, such as friction, of the forces between particles).  In our sketch this is achieved for a density $\phi_J$, which in general will be preparation dependent.  When the isostatic state reached is disordered the system is said to be jammed, and the onset of mechanical stability at $\phi_J$ is the jamming transition.  The figure shows that for densities beyond $\phi_J$ particles would in general overlap; such configurations would be allowed in systems of soft or deformable particles only.  Mechanical characteristics, such as the spectral properties of vibrations, change in a singular manner across the jamming point.  The jamming transition is in principle a zero temperature phenomenon, and its relation to the thermal glass transition is an important question in the field.

\section{Dynamic heterogeneity and facilitation}
A long-standing puzzle in glass physics was the apparent similarity between high-temperature liquids and supercooled ones. What are the features that make the latter different from the former, besides the exceedingly slow dynamics? This is clearly a central question, whose answer can bring us closer to discover the fundamental physical mechanism 
inducing the glass transition. One of the main achievements of the last years was the 
discovery and the characterization of such a feature: the phenomenon called dynamical heterogeneity \cite{Ediger2000,Glotzer2000,Andersen2005,bookDH}. High temperature liquids are homogenous in space and time: there is no essential difference 
in the way particles move in different regions of the liquids, nor there is a difference in the way a given particle moves now and, say, a fraction of relaxation times later. Supercooled liquids are not like that. They are characterized by spatial clusters of fast and slow moving particles. Moreover, a given particle can remain slow for a certain time and then become fast later, displaying intermittent behavior. This phenomenon, initially found in supercooled liquids, was later shown to be common to many other glassy liquids from colloids to granular media \cite{Dauchot2005,Keys2007,Mayer2004,bookDH}. Recently, it has been also found in active matter \cite{Angelini2011} and suggested to take place for several quantum systems \cite{garrahanquantum,nussinov}.
In the following, we shall discuss the four hallmarks of dynamics heterogeneity.

\noindent
{\em Non-exponential relaxation in time and large distribution of timescales}. Time-dependent equilibrium correlation functions of supercooled liquids, such as coherent and incoherent dynamical structure factors, show a time-dependence which is slower than exponential and well described by a stretched exponential. Correspondingly, linear responses, such as dielectric 
susceptibility, exhibit a non-Debye behavior characterized by a broad loss peak. All these behaviors are manifestations of the very same phenomenon: 
a large distribution of local relaxation times. One of the most striking consequences is the so called violation of the Stokes-Einstein relation \cite{Schweizer2007,Berthier2011}, {\it i.e.} the fact that 
the self-diffusion coefficient $D_s$ decreases not as fast as the 
viscosity $\eta$ increases, contrary to what happens in a homogeneous liquid.
The product $D_s \eta$ indeed increases by 2-3 orders of magnitude approaching the glass transition. Physically, this violation means that two different measures 
of the relaxation time $R^2/D_s$ and $\eta R^3/T$ ($R$ is the inter-particle distance)
do not lead to the same timescale up to a constant factor---a strong hint of the existence of a broad distribution of relaxation timescales \cite{Hodgdon1993,Tarjus1995}.

\noindent
{\em Dynamical correlations.} The way in which a supercooled liquid relaxes is not homogeneous as shown in the Fig. 1(ii). One clearly sees that the average behavior is not representative of the typical one: some regions are faster than the average,
others are slower. The more the liquid is supercooled, the larger are the slow and fast  clusters intervening during the relaxation process. From a statistical point of view, this phenomenon can be captured by studying dynamical correlations measuring 
to what extent the local relaxation taking place during the interval of time $0,t$ is correlated to local relaxation processes at a distance $\ell$ away from it, that also take place within $0,t$. Since local relaxation is probed by two point functions, such as density-density for instance, dynamical correlations are obtained by four point functions. A lot of studies have been devoted to the characterization of four point functions, in particular the so called dynamic susceptibility $\chi_4$, to the extraction of a corresponding dynamic correlation length and also to 
the analysis of the geometry of the fast and slow clusters (compact, fractal, string-like) \cite{Tarjus2011,Harrowell2011}.
Numerical simulations were instrumental for these studies since one can track the positions
of all particles as a function of time and, hence, potentially measure all kinds of observables. The situation is different in experiments, where measuring four point functions 
is instead much more challenging, at least for molecular liquids. In consequence, alternative methods to probe the number of correlated molecules $N_{corr}$ have been developed. It has been proposed, and found experimentally, that the growth non-linear susceptibilities 
is related to the one of $N_{corr}$. Alternative experimental estimates 
based on exact lower bounds were also proposed and applied \cite{BBBJ2011,Richert2011}. All that have provided an entire new sets of inputs and constraints for theorists, as it will be discussed later on.

\noindent
{\em Dynamic facilitation.}
Looking again Fig.\ 1(ii), one can find another facet of dynamical heterogeneity: dynamic facilitation \cite{Chandler2010}. This is the property by which a local region which undergoes relaxational motion in a supercooled liquid, or in similar slow relaxing material, gives rise or {\em facilitates} a neighbouring local region to subsequently move and relax \cite{harrowell,Garrahan2002,Chandler2010}.  This in turn leads to spatial segregation of relaxation and thus naturally to dynamic heterogeneity.  Dynamical facilitation is the key property of kinetically constrained models of glasses \cite{Ritort2003} on whose detailed study DF theory is based. Recently, it has been also taken into account within RFOT theory \cite{wolynesdf}, but it remains a by-product and not the key ingredient in this context.
Dynamical facilitation becomes explicitly evident  
in the trajectory movies from which the time frame of Fig.\ 1(ii) is taken; see for example the embedded media in Ref.\ \cite{Keys2011} which shows that heterogeneous growth of the relaxed clusters of molecules coloured dark red in the blue background of unrelaxed molecules.  Proving that facilitation is indeed the dominant mechanism in a generic glass former---that is, that a local relaxation has a very high probability of happening nearby another local relaxation after a certain time, which is short compared to the macroscopic relaxation time but large compared to the microscopic one, giving rise to propagation of mobility---is difficult since one has to disentangle motion which does not lead to relaxation (e.g. rattling and local vibrations) from the one that effective does so \cite{Elmatad2012}. This coarse-graining procedure was performed independently and in different ways in two recent numerical simulations of supercooled liquids and also in experiments of granular glasses \cite{harrowelldauchot,Keys2011,CDB}. It was shown that facilitation indeed takes place and account for a substantial part of the global relaxation. The insight of DF theory \cite{Garrahan2002,Chandler2010} is that facilitated relaxation becomes increasingly the dominant mechanism for global relaxation 
when lowering the temperature and approaching the glass transition, and that other means of motion, i.e.\ local relaxations not induced by facilitation, do not play a substantial role in this regime.  This assumption, that posits the origin of slow dynamics in facilitation and a decreasing density of local facilitation events with temperature, remains to be directly tested in experiments.

\noindent
{\em Propensity for relaxation and soft modes.}
A related issue is the relationship between dynamic heterogeneity and static 
properties. Is there something in the structure that is the cause of dynamic heterogeneity? A partial answer to this question was provided by numerical simulations correlating propensity maps to normal modes analysis. Propensity was introduced to measure how the likeliness of local motion depends on the structure \cite{propensity}: given an initial  configuration one studies all possible dynamical evolutions generated by sampling the initial velocities from the Maxwell-Boltzmann distribution. It was shown that propensity is large in regions where low frequency normal modes have a high amplitude \cite{harrowellcooper,harrowellreichman}.  
By summing over soft local modes one can identify the regions where the dynamics is likely to take place \cite{harrowelldauchot} and, hence, partially explain dynamic heterogeneity. See also \cite{wyart} where similar results were obtained for out of equilibrium hard spheres.  An alternative interpretation, based on the study of certain kinetic constrained lattice gases \cite{Ashton2009}, is that these correlations are simply a consequence of the presence of the localised excitations that facilitate dynamics and simultaneously disrupt the elastic network giving rise to an excess of soft vibrational modes. 
A full understanding of the role of soft-modes, especially at lower temperature, closer to the glass transition, is still lacking and points of view diverge on the most likely outcome, as we shall discuss later.

\section{Search for Static correlations}
As discussed in the introduction, there is no obvious structural change accompanying  
the slowdown of the dynamics of supercooled liquids: glasses look like liquids that have stopped to flow. 
On the other hand, one cannot exclude {\it apriori} the existence of subtle static correlations which are not picked up by simple correlation functions investigated so far---the search for locally preferred structures has indeed a long history in the field of glass transition (with a mild 
success) \cite{Charbonneautarjus,royal,tanaka}. The main difficulty is that one does not know what is looking for:
one can recognize by eye that a crystalline or quasi-crystalline solid is ordered but how to make the difference between 
an amorphous ordered structure from a completely disordered one?  

An important recent step was to come up with a precise definition of static order in an amorphous structure. This has lead to a flourishing of proposals for static lengths whose mutual relationship is not fully clear yet. The main idea, common to all these proposals, is that an ordered structure is characterized by some kind of predictability or, equivalently, 
by low entropy \cite{levinekurchan,cammarotaPR}: knowing part of it allows one to predict the rest with good accuracy since {\it the structure is ordered}. 
Instead, in a completely disordered structure, knowing a part gives no information about the rest. In order to unveil the existence of amorphous order, the procedure that was
introduced originally consists in pinning particles from an equilibrium configuration and studying the effect induced on the remaining free ones: if the number of equilibrium configurations sampled by the remaining free particles is strongly reduced compared to the unpinned case then the original system is indeed characterized by amorphous order \cite{Bouchaud2004}. The set of pinned particles considered originally is a region outside a spherical cavity. The question, in this case, is whether this particularly boundary condition reduces the sampled configurations just to small fluctuations around a given amorphous structure. Technically, one measures 
the overlap at the center of the cavity between two equilibrium configurations which are identical outside the cavity.
An overlap that remains high up to cavity-radii of length $\ell$ proves the existence of static order up to this length-scale. This procedure is very similar to the one used, e.g., for the Ising model: setting the boundary spins up at low temperature forces all configurations 
to be in the up-state. The crucial difficulty in supercooled liquids is that we do not know apriori what is the correct pinning boundary condition; 
the trick is to use the fact that instead the system ``does'' it, if it is indeed ordering. This is the 
reason why the configuration from which particles are pinned is an equilibrium one.  
This procedure was first implemented numerically in \cite{CGV,BBCGV} and lead to the first 
proof that static amorphous order grows (mildly) approaching the glass transition. Several other numerical investigations have followed and other arrangements of pinned 
particles have been studied (such as wall or sandwich geometries) \cite{refsgeometries1,refsgeometries2,refsgeometries3,refsgeometries4}. 
These works were  
instrumental and set the stage for
a thorough investigation of static amorphous order in supercooled liquids. 

The crucial questions that remain to be addressed now are: Is the growth of the static correlation length a cause or a pure by-product of the increase of the relaxation time? 
What is the relationship between dynamic and static correlation lengths? Are all static
correlation lengths essentially equivalent or do they capture different physical effects? 
As for now, we only have partial answers; for instance, it was proven that a diverging relaxation time implies a diverging relaxation length \cite{montanarisemerjian}. 
Numerical simulations have shown that dynamic correlations extends over a range larger than static lengths \cite{dynstat,dynstat2}, as expected theoretically at least in the regime one can focus on in numerics \cite{Garrahan2011,Franz2011}.
A direct analysis of the behavior of the static length was also performed in a few models of glassy dynamics belonging to RFOT and DF theories: in the Kac limit \cite{franzkac} and starting from a Ginzburg-Landau action \cite{wolynesschmalian} for the former 
and in the plaquette model \cite{jackgarrahan} for the latter.   
We expect that the efforts for obtaining complete answers to those questions will concretize in a future major research program, mirroring the one on dynamical correlations, that should allow us to fully comprehend the role of long-range amorphous order for the glass transition and, hence, to what extent the transition is related to a static critical phenomenon.

\section{Relationship with jamming}
Glass transition and glassy behavior can be found in systems that are microscopically very different from supercooled liquids. A notable example are colloidal particles which interact through a steep repulsive potential \cite{pusey}. These are well modeled as hard spheres and they display liquid, crystal and glass phases. In this case the control parameter driving the transitions
is the density (or volume fraction) and thermal motion only affects the value of the short-time diffusion constant. Typical length and time scales are very different from molecular systems: particle size and inter-particle distance are of the order of $\mu m$ instead of fraction of 
$nm$, whereas the collision time is of the order of $ms$ instead of $ps$. 
Remarkably, despite these microscopic differences, their dynamics is glassy in a way very similar to the one of supercooled liquids. Another kind of systems that has been recently studied from this perspective are granular glasses: assemblies of grains that form crystals and glasses when their volume fraction is large enough. Not only these systems are characterized by even more different scales---millimeters instead of nanometers, seconds instead of picoseconds---but moreover their glassy behavior arise out of equilibrium since grains need mechanical forcing in order to move: in this case the glass transition is due to the slow down of the dynamics of the out of equilibrium steady state reached thanks to the external drive. These differences notwithstanding, there is a clear similarity between their glassy dynamics and the one of supercooled liquids \cite{Dauchot2005,Keys2007,dauchot}. All these findings suggest that glassy behavior and the glass transition have a high degree of universality and their study encompasses a very wide spectrum of physical systems.

Well before discovering that colloids and grains can form glasses, it was known that they can jam when rapidly compressed: the most famous example is the randomly close-packed state originally proposed and studied by Bernal \cite{bernal} (see \cite{torquato} for a more recent perspective). In recent years the parallel and independent research effort on jamming has started to join forces with the one on the glass transition. A natural question that came up is to what extent the jamming transition of colloids and
granular media is related to their glass transition.  In Ref.\ \cite{liunagel} it was suggested that they are two facets of the same story. This was vividly proposed introducing the jamming phase diagram, where the critical point associated to the jamming transition, called point J, governs the whole slow dynamics behavior of liquids, grains and granular media subjected to thermal noise and drive.  This provocative idea triggered a lot of research activity. By now, it is clear that the situation is more complicated 
than originally thought. First, point-J is not a point but a line since the density at which a system jams depends on the protocol used to compress it \cite{berthierpointjs}. Moreover, fast compression makes the system explore regions of the configuration space different from the ones sampled at equilibrium. In consequence, jamming and glass transitions are related to quite different physical phenomena, as first suggested in \cite{krzakalakurchan} and then  proven numerically for three dimensional hard spheres \cite{berthierwitten,berthiersollich}. This notwithstanding, the study of the J-point has provided a new interesting reference frame to think about disordered and glassy systems. The jamming transition of frictionless particles is related to the existence of soft modes: it was shown that an assembly of soft elastic spheres loose its rigidity because low energy modes become very soft and eventually unstable by decreasing the density below point-J. 
The presence of these soft modes makes the physical properties of disordered jammed systems very different from the one of usual elastic solids \cite{Liu2010,procaccia}. 
Understanding the role of modes in the slow dynamics has been a long-standing {\it leit motif} in the field of the glass transition. The soft-modes appearing at the J-point provide a new way to think about that \cite{Liu2010}.   
As usual in the glass transition problem, which is an intermediate coupling problem,
this is a paradigm that serves as starting point on which one endeavors to construct a
more realistic theory. It has been certainly useful to improve our understanding of glassy and
disordered systems. Whether as starting point it is close enough to the correct theory to be useful only the future will tell. The crucial point is how much the glassy dynamical behavior  
can be described in terms of modes: recent numerical simulations have shown that this is the case for moderate super-cooling \cite{harrowellreichman,wyart} as it was expected qualitatively at least from certain theoretical perspectives, such as RFOT theory \cite{RFOTLW,RFOTBB}. 
Whether this remain true at lower temperatures or higher volume fractions closer to the glass transition, where the dynamics is known to be activated and super-Arrhenius is an important open question that hopefully will be settled in the future. From the theoretical point of view 
it is hard to explain activation in terms of modes; several approaches aimed at describing the glass transition have been developed further to take into account the possibility of soft regions as discussed previously, but this new ingredient is viewed only as a byproduct and not  
as not the main cause for activated dynamics.

\section{Recent Theoretical advances}

The complexity of the glass transition problem is underlined by the fact that a large number of disparate theoretical proposals have been put forward over the years to explain it.  As we have emphasised above, the main recent developments have centred around real space phenomena, such as dynamic heterogeneity, transport decoupling, the search for accompanying structural correlations, and the relevance of fluctuations near structural rigidity or jamming.  Most competing glass theories have responded to these developments and evolved accordingly. For example, in the case of MCT, recent developments have seen extensions of what in origin was a mean-field theory for global correlators such as time-dependent structure factors or propagators that captured departure from standard liquid-state behaviour, into a fully fluctuating field theory within which multi-point correlations can be calculated and in which the MCT singularity acquires meaning as a critical point with concomitant critical properties for both time and spatial fluctuations, which in turn can then be directly contested against real space observations (and which would also allow systematic computation of the fluctuation induced effects that should signal departure from ideal MCT predictions); see e.g.\ Refs.\ \cite{IMCT,JCPlong,fluctpapersparisi}. 

In the context of RFOT, recent works have aimed at studying thoroughly its critical properties and real space behavior, performing the first renormalization group analysis \cite{MKFT,dysonrem} and the first numerical investigations of the critical exponents $\theta,\psi$ \cite{cavagna,sastry}, whose values were originally argued to be equal to $3/2$ in three dimensions \cite{KTW} (numerical results do not seem to confirm these expectations though). One other significant development, mentioned above, has been that of the mean-field theory of hard sphere glasses (see here for a detailed review \cite{parisizamponi}), which has provided clear and quantitative thermodynamic predictions for hard-spheres which should be valid in high dimensions, and has allowed the explicit unification of glass transition and jamming problems as understood within an RFOT scheme.  It is worth remarking that an important aspect of this work is that is ``simply'' about hard spheres.  This allows to counter the argument that RFOT is essentially a ``landscape'' theory, i.e.\ one requiring an explicit complex energy landscape, while for liquids the paradigm is hard spheres 
(as, after all, modern liquid state theories such as WCA \cite{Weeks1971} are built around hard sphere reference systems) where what matters is the metric structure of their configuration space. 
Although some predictions of the RFOT approach to hard sphere glasses are in 
striking agreement with simulations, e.g. some features of the pair distribution function, 
others are not, in particular the critical exponents of the jamming transition \cite{ikedaberthierbiroli}. This is quite puzzling since these exponents are not expected to depend
on dimension and, hence, should be correctly accounted for by mean-field theory. 
This is a hint that the construction of the solution of hard sphere glasses in the limit of infinite dimensions, and hence, of a full-fledged mean-field theory has still to be completed. The recent advances presented in \cite{kurchanparisizamponi} suggest that this goal is within reach.  A similar research effort aiming at making mean-field theory quantitative was developed in the context of MCT, which has been generalized to cope with non-equilibrium steady states such as sheared glassy liquids, a very relevant physical situation \cite{fuchscates,reichmanmiyazaki} and in the context of RFOT to take into account fluctuations
of the mosaic state \cite{Lubchenko2007,wolyneswalcwack}. 
  
As we discussed above, dynamic heterogeneity emerges naturally within DF theory.  A central development within this approach \cite{Jung2004,Berthier2005} has been a way to understand transport decoupling, such as the breakdown of the liquid state Stokes-Einstein relation between viscosity and self-diffusion rate \cite{Schweizer2007}, as a direct consequence of fluctuations in the dynamics.  The key observation has been that 
dynamic heterogeneity implies intermittency in the waiting times between local events leading to relaxation due to a distribution of timescales that affect different transport processes differently, depending on the fundamental length scale in play.  This analysis can be boiled down to the existence of two typical timescales, sometimes referred to {\em persistence} and local {\em exchange} times \cite{Jung2004}.  The persistence time is the typical waiting time for a local relaxation event, such as a molecule moving irreversibly by a distance comparable to its size, to happen for the first time.  The local exchange time is the typical waiting time for such an event to occur again once it has already happened.  The former dominates structural relaxation, while the latter dominates diffusion.  In an intermittent system these two timescales can be very different due to, borrowing terminology from quantum optics \cite{Barkai2004}, event ``bunching'': facilitation implies correlation both in space and time of relaxation events, so that when one happens many others typically follow.  This is the essence of decoupling from the DF point of view.  A simple scheme to compute time correlations, both two-point such as intermediate scattering functions or multi-point such as four-point susceptibilities, can be constructed \cite{Berthier2005} in terms of a continuous-time random walk (CTRW) \cite{Montroll1965} for probe molecules, where the effect of the rest of the system is encoded in the distribution of waiting times for the fluctuating walk.  Just like other aspects of DF this description of decoupling has emerged from the study of idealised kinetically constrained models \cite{Jung2004,Berthier2005}, but has then shown to be applicable to more realistic systems \cite{Hedges2007,Chaudhuri2007}.  Similar ideas of persistence/exchange and CTRW have also been extended to dynamics in terms of transitions between coarse-grained minima or metabasins \cite{Heuer2008}. 

Alternative explanations for decoupling have been developed within RFOT.  In the MCT regime Stokes-Einstein violation 
is expected only below the upper critical dimension $d_u=8$ and is due to critical fluctuations \cite{BBSE}. In the activated regime, decoupling is due to the local fluctuations of 
configurational entropy that induce a distribution of relaxation times \cite{XiaWolynes}.
The MCT prediction on dimensional dependence has been recently tested in numerics by simulating hard spheres from three to nine dimensions \cite{charbonneau}: the results confirm that above eight dimension decoupling disappears. Actually, using space dimensionality as a varying control parameter to test glass theories is a protocol 
that started to attract a lot of attention and appears to be 
a promising research avenue for the future \cite{charbonneau,MCThighD1,MCThighD2}.

A final recent set of results we would like to mention relate the frustration limited domain theory \cite{Tarjus2005}. Numerical simulations of liquids on hyperbolic planes have shown 
that indeed the competition between space curvature and the tendency to develop long-range order gives rise to defect formation and, consequently, to slow dynamics \cite{refsgeometries1}. This shows that the physical mechanism envisioned by this theory to explain the glass transition indeed is a viable one. Whether it is at work for three dimensional liquids in flat space remains an open question.  
  
Despite the many advances in the last few years the central questions in the glass field remain the same: what is the nature of the true transition behind the observed experimental kinetic glass transition; what drives this transition and how does it manifest in the observed phenomenology; and given that it is clearly elusive under normal conditions, are there any protocols under which it can be directly accessed.  Most if not all theories that have been put forward, be it MCT and all its variants, frustration-limited domains and similar theories based on more or less standard order/disorder, RFOT and DF theory,  posit in one way or another some sort of transition as the ultimate underlying cause for glass formation.  What varies significantly is the assumption on the nature 
of the transition---whether it is a finite temperature, a zero temperature, or an avoided phase transition, whether it is thermodynamic or purely dynamical---and the consequences of the fluctuations associated to it on the observed phenomenology.  This in turn is related to what different theories consider the fundamental excitation mechanisms that give rise to relaxation.

If the experimental glass transition is as many believe the consequence of some underlying singular phenomenon, it is clear that, whatever its origin, it is hard to access.  RFOT predicts such singularity is a thermodynamic phase transition at some temperature $T_K$ which is in principle inaccessible in experimental timescales as it is below the temperature supercooled liquids fall out of equilibrium.  For DF theory in turn, the underlying transition is one between dynamical phases, an equilibrium one for the liquid and a non-equilbrium one for the glass, but again the transition is in principle not accessible in normal dynamics which only manifests its proximity.  In order to overcome this inaccessibility problem new theoretical techniques have been developed recently, two of which we will discuss now.  Both amount to biasing ensembles, of configurations in one case in order to bring a thermodynamic RFOT transition, if it exists, to within reach; and of trajectories in the second case, in order to access the non-equilibrium order/disorder transition expected from DF theory.

\subsection{Biasing ensembles of trajectories: the $s$-ensemble}

DF has evolved as a theory from the study of idealised models of glasses, so-called kinetically constrained models (KCMs) \cite{Ritort2003,Garrahan2011}, such as facilitated spin lattice models.  These idealised systems capture many of the basic features of glassy relaxation, in particular dynamic heterogeneity, and they do so in the absence of any interesting or singular thermodynamics.  Furthermore, recent detailed studies of the dynamics of atomistic liquids suggest that the basic tenets of KCMs, such as localised non-interacting excitations, facilitation and hierarchical dynamics (specifically as in East like facilitated models \cite{Ritort2003,Garrahan2011}), are present in supercooled liquids \cite{Keys2011}.  Since in KCMs dynamic heterogeneity is the consequence of complex structure in dynamical trajectories, it is therefore natural to study the dynamics of glass formers from the point of view of a ``statistical mechanics of trajectories'' rather than configurations.  The natural framework to do this is provided by (dynamical) large-deviation theory \cite{LD,Lecomte2005}. 

Lets denote by $X_t$ a dynamical trajectory of a many-body system of total time extent $t$, i.e.\ the succession of configurations ${\cal C}$ of the system from some initial condition ${\cal C}_0$ all the way to a final one ${\cal C}_t$, $X_t \equiv (  {\cal C}_0, {\cal C}_{\delta t}, \ldots, {\cal C}_t)$.  For simplicity we consider equilibrium trajectories, as we wish to study the slow dynamics in the supercooled regime.  The dynamics generates a trajectory $X_t$ with a certain probability, $P[X_t]$, which defines the (unbiased) ensemble of trajectories.  Under supercooled conditions these trajectories display the space and time fluctuations that manifest in dynamic heterogeneity.  In order to characterise trajectories we define a trajectory observable which will serve as a (dynamical) order parameter.  The natural one for the glass transition problem is the {\em dynamical activity} \cite{Merolle2005,Garrahan2007,Hedges2009}, which we denote by $K$, defined as the total number of configuration changes in a trajectory for a system with discrete degrees of freedom, such as a lattice model, or a suitable coarse-graining in a continuous force system such as an atomistic liquid.  In practice, the precise nature of $K$ does not matter, as long as it is a time-integral, i.e.\ extensive in time, and it captures motion leading to structural relaxation \cite{royal,Speck2012,Fullerton2013}.  Highly relaxing trajectories will have $K$ large as there would be a lot of motion, while sluggish or arrested ones will have $K$ small.  In fact, if one considers a trajectory displaying dynamic heterogeneity, such as the one of Fig.\ 1(ii), activity appears to be spatially segregated, with space (and time) regions of high activity, coloured red in Fig.\ 1(ii), coexisting with those of low activity, blue in Fig.\ 1(ii).  This coexistence is suggestive of an underlying active-inactive transition in trajectories.

If $K$ is the order parameter then we should consider the corresponding order parameter distribution, $P_t(K) \equiv \sum_{X_t} P[X_t] \delta(K - K[X_t])$, where the average is taken over the whole ensemble of equilibrium trajectories.  The same information as in $P_t(K)$ is contained in the moment generating function, $Z_t(s) \equiv \sum_{X_t} P[X_t] e^{- s K[X_t]}$.  Note the role of $s$: the time-extensive order parameter $K$ is in the {\em count} of transitions between (coarse-grained) configurations, and $s$ is a ``counting'' field.  At long times, the generating function acquires a large-devation form, $Z_t(s) \approx e^{t \psi(s)}$, where the large-deviation function $\psi(s)$ is the generating function for cumulants of $K$ at long times \cite{Lecomte2005}.  
This framework amounts to a {\em thermodynamics of trajectories} \cite{Merolle2005,Garrahan2007,Hedges2009}: the relevant ensemble is the set of many-body trajectories of time extent $t$ (cf.\ many-body configurations in an equilibrium statistical ensemble), activity $K$ and counting field $s$ are the extensive observable and its intensive conjugate field (cf.\ magnetisation/magnetic field, or number of particles/chemical potential), the large size limit is given by $t \to \infty$ (cf. large volume limit), and in that case the ``partition sum'' $Z_t(s)$ is determined by the {\em dynamical free energy} $\psi(s)$ (cf.\ a Gibbs free-energy in the static magnetic case, of a grand potential in the grand canonical one).  In particular, just like a free-energy, the analytic structure of $\psi(s)$ as a function of $s$ determines the dynamical phase structure of the ensemble of trajectories, and a singularity in $s$ indicates a phase-transition in such ensemble.

\begin{figure*}[th]
\includegraphics[width=1.7\columnwidth]{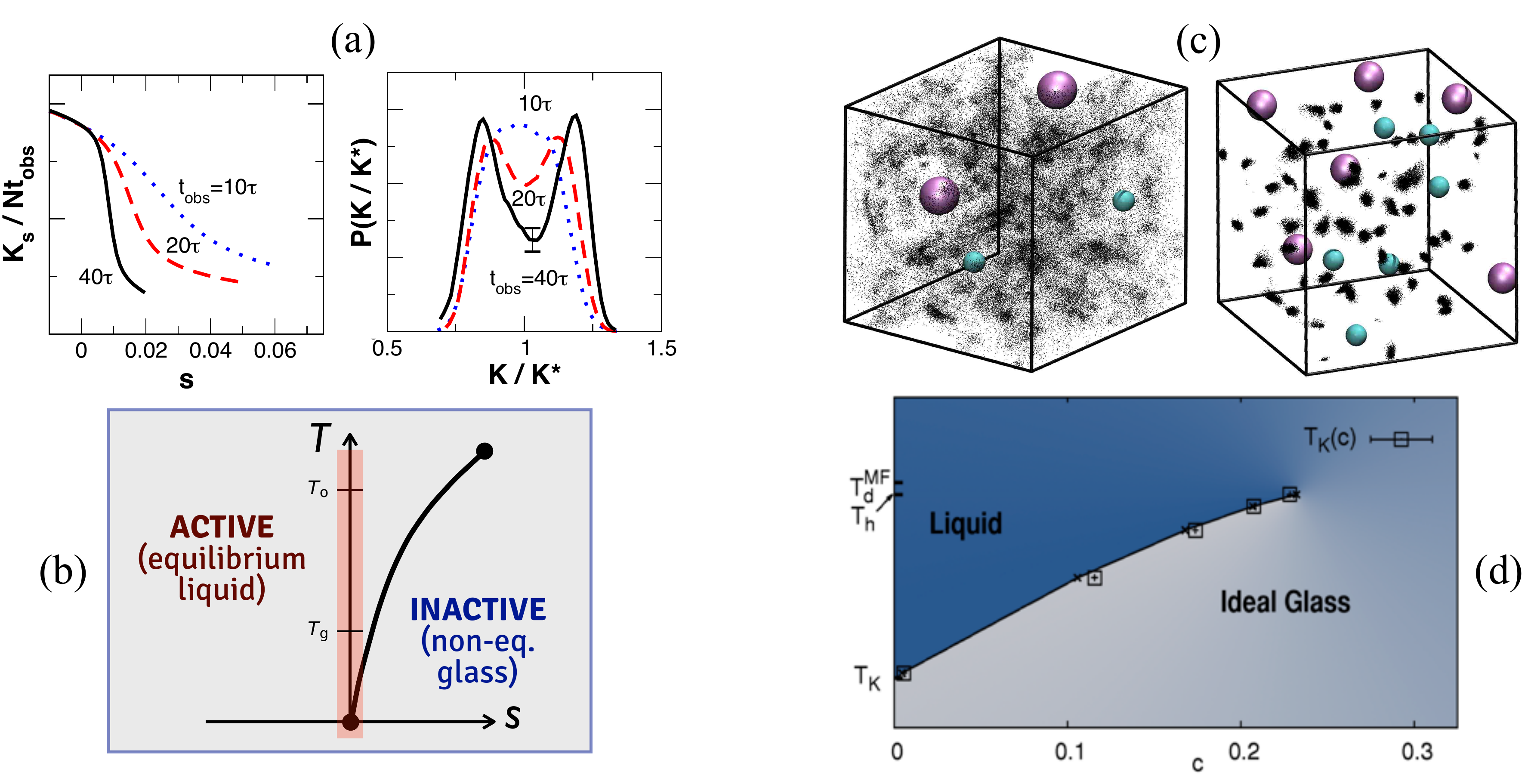}
\caption{Biasing of ensembles in order to access dynamical or thermodynamical glass transitions. (a) $s$-ensemble for a Lennard-Jones binary mixture, from Ref.\ \cite{Hedges2009}, biased by dynamical activity through the counting field $s$. The average activity $K_s$ displays a crossover from a large value at $s=0$ to a small one at $s>0$; this becomes sharper for longer observation times ($\tau$ is the alpha-relaxation time). At $s_c$ the order parameter distribution displays the bi-modality characteristic of a first-order transition, between a dynamically active equilibrium phase, and a dynamically inactive non-equilibrium one.  (b) Suggested ``space-time'' phase diagram: while dynamics takes place within the active phase, the closeness of the first-order transition to the inactive phase gives rise to fluctuation behaviour manifested in dynamic heterogeneity.  (c) Result of numerical simulation of pinned harmonic spheres \cite{kobberthierpinning}. Large spheres represent pinned particles (rescaled in size by a factor 0.5), small dots are the superposition of the positions of fluid particles obtained from a large number of independent equilibrium configurations in presence of the pinned particles. (d) Phase diagram for pinned particles obtained by renormalization group analysis in \cite{PNAS} starting from a Ginzburg-Landau action.  
\label{Fig2}}
\end{figure*}

For certain models $\psi(s)$ can be computed analytically via transfer matrix techniques \cite{Garrahan2007}.  More generally, $\psi(s)$ encodes a biased ensemble of trajectories, $P_s[X_t] \propto P[X_t] e^{- s K[X_t]}$, where trajectories with smaller/larger activity are enhanced/suppressed as compared to the standard dynamical ensemble, depending on the value of $s$. 
This is the so called $s$-ensemble \cite{Hedges2009}, which can be explored numerically via path sampling techniques.  A sudden change in the properties of the $s$-ensemble is indicative of a phase-transition in the space of trajectories.  
In Fig.\ 2(a) we show such computation of the $s$-ensemble \cite{Hedges2009} for a Lennard-Jones binary mixture, the standard atomistic model of a glass former.  The activity changes from a higher, i.e.\ more active value, to a lower, i.e.\ less active, as $s$ increases, as expected.  What is notable is that the change becomes sharp with increasing observation time, and that the value $s_c$ where this takes places is close to $s=0$, the point at which (unbiased) dynamics occurs.  This is reminiscent of the jump of an order parameter at a first-order transition.  And indeed at $s_c$ the order parameter distribution becomes increasingly bi-modal with increasing observation time.  

The implication is that equilibrium supercooled liquid dynamics is taking place very close to a first-order dynamical (or ``space-time'') transition to a non-equilibrium inactive phase.  The mesoscopic fluctuations due to the proximity to this phase boundary manifest in the dynamical heterogeneity of the equilibrium relaxational dynamics.  Results from KCMs \cite{Elmatad2010} and from atomistic simulations, would suggest that the ``space-time'' phase diagram looks like the one of Fig.\ 2(b), with a first-order phase boundary which gets closer to the point of actual dynamics $s=0$ with decreasing temperature (and which may end at a dynamical critical point at temperatures above that of the onset of heterogeneous dynamics).  For temperatures below the onset of heterogenous dynamics, equilibrium dynamics (at $s=0$) takes place close to the transition line to the inactive phase giving rise to mesoscopic fluctuations---``space-time bubbles'' of inactivity \cite{Chandler2010}---that manifest as dynamic heterogeneity.

\subsection{Biasing ensembles of configurations: random pinning}

The RFOT theory predicts a finite temperature thermodynamic transition at which 
amorphous long-range order sets in. Correspondingly, time and length-scales diverge 
following an activated dynamic scaling: the typical length-scale (the point-to-set) 
diverges as a power law, $\xi\propto (T-T_K)^{-1/(d-\theta)}$, and the time-scale 
in an exponential way, $\tau\propto \exp((T-T_K)^{-\psi/(d-\theta)})$ \cite{RFOTBB}. There is an inherent 
difficulty in testing such a critical phenomenon: the time-scale increases so fast that 
before approaching $T_K$ the system inevitably falls out of equilibrium.  Thus, analyzing the critical properties or even showing that there is a phase transition may seem out of reach.
A way to short-circuit this difficulty was proposed recently in \cite{PNAS,PNASlong} and it can be considered a counterpart, in the RFOT context, of the procedure outlined above for DF theory. The main idea is that by pinning a fraction $c$ of particles at random from an equilibrium configuration one can induce a glass transition even at rather high temperature, see the phase diagram in panel (d) of Fig.2. One of the great advantages of this procedure is that just after pinning the unpinned particles are already at equilibrium and, hence, Ê
one can access the ideal glassÊphase easily and be able to approach the glass transitionÊfrom both sides; in consequence 
showing the existence of the transition and probing its critical properties becomes possible. 
This places the problem of the glass transition on a similar footing of other transitions
characterized by activated and very slow dynamics, such as the Random Field Ising model transition for which the numerical analysis has proven to be very challenging but conclusive results have been obtained.
 
RFOT theory is based on the assumption that there are exponentially many amorphous glass phases in which a supercooled liquid can freeze. supercooled liquid become very viscous but do not freeze precisely because they have too many choices for doing that: the tendency to lower the free energy by 
ordering in a given amorphous phase is compensated by the gain obtained by disrupting 
the long-range order and sampling all the possible different phases. This competition leads to the so called mosaic state which is a kind of micro-phase separated phase in which the number of possible phases is actually huge. By approaching $T_K$, the configurational entropy decreases, {\it i.e.} the number of available glass phases diminishes, and the system
eventually orders at $T=T_K$ \cite{RFOTLW,RFOTBB}. By pinning a fraction $c$ of particles from an equilibrium configuration at temperature $T$, one biases the configurations sampled by the system and decreases the number of available glass phases: only the ones compatible with the positions of the pinned particles survive. Thus, the configurational entropy decreases when increasing $c$ and, within RFOT, it is expected to vanish at a given $c_K(T)$, hence inducing a glass transition. Physically, for $c<c_K$ the biasing field induced by the pinned particles is not large enough to freeze the system in a given amorphous phase, whereas it is instead able to do so for $c>c_K$, see panel (c) of Fig. 2 which is the result of the numerical simulations presented in \cite{kobberthierpinning}. The order parameter of the transition is the overlap between the initial reference configuration and an equilibrium one obtained in presence of the pinned particles. 
At $c=c_K$ the overlap displays a discontinuous jump and is characterized by a bi-modal distribution very similar to the one obtained within the {\it s}-ensemble (panel (a) of Fig.2) 
The glass transition line $c_K(T)$ ends in a critical point where the glass transition becomes continuous and falls in the universality class of the Random Field Ising model \cite{PNAS,PNASlong} (for a thorough analysis of the phase diagram and the transition lines see \cite{cammarotaepl}). 

The protocol of pinning particles opens the way to new investigations of the glass transition. First, it allows one to test the glass transition also coming from the glass side, an impossible task in the usual case. Second, it is a direct test of whether the mechanism at the root of RFOT plays indeed a role, since pinning is not expected to induce a glass transition within other theoretical frameworks, in particular DF theory \cite{jackberthier}. Finally, by extrapolating results obtained by pinning above $T_K$ one could obtain information on what happens at $c=0$, i.e. on whether the $c=0$ endpoint of the random pinning glass transition line pinches the $y$-axis at a non-zero temperature, corresponding to a finite temperature RFOT. Recent simulations from atomistic simulations obtained the first evidences of the random pinning glass transition \cite{kobberthierpinning} (see also \cite{kim1,kim2,karmakarprocaccia}) and confirmed some theoretical predictions \cite{PNAS}.

We should point out yet another way of biasing that is relevant to study the glass transition. It was introduced in Ref.\ \cite{FP} and is called $\epsilon$-coupling: 
It consists in introducing a bias in the thermodynamics of a glassy system to favor configurations correlated with a reference one, chosen from the equilibrium measure at temperature $T$. The bias is introduced by adding to the Hamiltonian an external attractive potential favoring density profiles similar to the one of the reference configuration. Although at first sight similar to the random pinning procedure, it leads to a standard first order phase transition: for a critical value $\epsilon_c(T)$, which is temperature dependent and vanishes for $T\rightarrow T_K$ the system has a discontinuous transition between two phases, with low and high overlap with the reference configuration. Although the transition at $\epsilon_c(T)$ is not a glass transition (it becomes so for $\epsilon=0$ {\it only}), theoretically
it is expected to be a consequence of the one taking place at $T=T_K$ and $\epsilon=0$ and, hence, provides valuable information on it. The numerical work performed in \cite{cavagnacammarota} presented evidences that the $\epsilon$-coupling indeed leads to a first order phase transition. 

In conclusion theoretical and numerical works have unveiled by biasing the thermodynamic measure that supercooled liquids appears to be in proximity of thermodynamic phase transitions which are direct consequences of the competition between configurational entropy, associated to a large number of metastable states, and free energy of those states. A thorough study of these transitions and their relationships with the behavior of unbiased liquids will certainly be a major research theme for the future.

%---FROM HERE---

\section{Outlook}

Here we have provided a brief overview of the current state of the glass transition field. 
Most recent developments have centred around issues of real space fluctuations and correlations, both in the dynamics and the statics, and this determined the main focus of the article.  From the numerous recent theoretical advances we have highlighted two proposed methodologies to gain access to the otherwise inaccessible ideal (dynamical or statical) transitions behind glass formation, the $s$-ensemble biasing of trajectories, and the biasing of configurations by particle pinning.  While fundamentally different at their outset, these two approaches have certain conceptual similarities.  Both posit the existence of an ideal phase that would correspond to the glass, a non-equilibrium inactive dynamical phase in the case of DF, or an ideal thermodynamic glass phase in the case of RFOT; under normal conditions, e.g.\ by cooling, the transition to this phase is difficult or impossible; but the phase boundary may be actually closer than it appears, which is revealed by considering an extended parameter space, see Figs.\ 2(b) and (d), where the extra parameter is the strength of the biasing.  At equilibrium supercooled conditions fluctuations related to this nearby inactive/glass/symmetry-broken phase will greatly determine the observed dynamics, giving rise to dynamic heterogeneity, a structural mosaic state, or variants depending on the theoretical perspective.  
Despite the differences (one is dynamical, the other thermodynamical; in one the transition line hits the temperature axis at $T=0$, while in the other it does so at $T_K >0$; and so on), the similarity between the phase diagrams of Figs.\ 2(b) and (d) is intriguing, and makes one wonder whether this may provide an avenue to find connections between different glass theories.

Needless to say, these theoretical approaches which we consider interesting need to connect to experiments.  While they seem well suited for theoretical or computational studies, it will be necessary to devise clear protocols for them to become experimentally testable.  For particle pinning it is not difficult to imagine an experimental situation where it is implemented on colloids held in place by optical tweezers \cite{PNAS}.  In the case of the $s$-ensemble the study of short time high-order dynamical cumulants \cite{Flindt2013}, also in principle accessible to experiments, may be the way to go.

From the experimental side, one of the most significant recent developments has been the discovery of {\em ultrastable glasses} \cite{Swallen2007,Swallen2009,Sepulveda2013}.
These are systems with a kinetic stability that is orders of magnitude beyond what can be reached by standard methods.  They are prepared \cite{Swallen2007,Swallen2009,Sepulveda2013} by vapour deposition in a way that avoids the kinetic trapping inherent in cooling protocols.  It is natural to think that these stable glasses could be directly related to either the inactive state of DF theory or to the ideal glass state of RFOT, so maybe experimentalists have already found a way to access these difficult to reach phases.  Either way, it seems evident that the more experiments clarify the preparation of these ultrastable materials and the relaxation out of them the more we will understand of the precise nature of the glass transition. 

Most of this review has dealt with properties of glass formers as the glass transition is approached from the equilibrium side, that is, in the case of liquids, in the supercooled regime.  The reason is that if the equilibrium dynamics is collective and complex enough to make a thorough understanding of it difficult, as it is in these systems, the out-of-equilibrium dynamics where one also has to consider preparation history is likely to be even more difficult to elucidate.  Nevertheless, number three of our stylised facts of the glass transition indicates that a key feature is how these materials respond to being taken out of equilibrium, and behaviour here is intimately related to what are the fundamental mechanisms behind the glass transition.  One such situation is that of {\em aging} \cite{Angell2000}, the very slow drift of either one-time or two-time properties of a glassy system.  This is a regime that has been studied thoroughly from the mean-field point of view \cite{Cugliandolo2002,Corberi2011}. It would be worth developing research aimed at going beyond mean-field theories, as it is already the case for equilibrium. Indeed, one should not consider aging as a limitation that prevents a direct study of the glass transition, but as an {\it atout} to probe the 
fundamental mechanism leading to glassy behavior from a different angle. For example, experiments on aging glasses \cite{Brun2013} may allow to determine the scaling relations between spatial correlations and timescales predicted by theory. Aging can also provide a new way to test the mosaic structure advocated within RFOT \cite{wolynesaging}. 
Similarly, it is possible to argue from the DF perspective that the same mechanisms that give rise to dynamic heterogeneity in equilibrium are responsible for the behaviour of non-equilibrium responses in standard calorimetry experiments \cite{Keys2013}.  We expect to see further developments in our understanding of the glassy out-of-equilibrium regime in the near future. 

We end by mentioning an area which is rapidly developing, that of {\em quantum glasses}.  
The last couple of decades have seen a revolution in the experimental realisation of quantum systems \cite{Haroche2006}. Experimental advances have allowed an unprecedented degree of control over ultracold gases \cite{Bloch2008}, trapped ions \cite{Liebfried2003}, superconducting circuits \cite{Schoelkopf2008}, and nano-electromechanical systems \cite{LaHaye2009}. Ultracold atomic gases are nowadays routinely created and used for the study of complex many-body phenomena such as quantum phase transitions, shedding light on open problems in condensed-matter physics \cite{Bloch2008}. 
This has brought to the forefront aspects of real-time dynamics of quantum many-body systems that directly connect to the glass transition problem.  Quantum glasses are of direct relevance to issues such as supersolidity \cite{Prokofiev,Biroli08,davis}, quantum annealing \cite{Das08,reviewMF}, aging in electronic systems \cite{Amir09}, thermalization \cite{Polkovnikov2011}
and many-body localization \cite{Altshuler,WolynesMB,Pal2010}.  Recent work has highlighted the interplay between classical and quantum fluctuations \cite{Markland2011}, argued how constrained dynamics may be relevant for quantum glasses \cite{Chamon2005,garrahanquantum}, discussed the possibility of the quantum analog of dynamical heterogeneity \cite{Nussinov2008,garrahanquantum,nussinov}, and shown the emergence of glassy-like dynamics in clean bosonic systems \cite{Poletti2012}.  But despite these advances, there is still much scope for ideas and methods of classical glassy systems to cross over into the quantum case.  It is safe to predict that the field of quantum glasses will make quick progress in the coming years.

\section*{Acknowledgements}

The ideas we presented here have been influenced and moulded by the interaction over many years with our colleagues and friends in the field.  In particular we wish to thank our recent collaborators on work we reviewed here.  These include: 
L. Berthier, J.-P. Bouchaud, C. Cammarota, R. Candelier, A. Cavagna, D. Chandler, O. Dauchot, Y.S. Elmatad, S.C. Glotzer, T.S. Grigera,  L.O. Hedges, D. L'Hote
A. Ikeda, R.L. Jack, Y. Jung, A.S. Keys, F. Ladieu, V. Lecomte, K. Miyazaki, D.R. Reichman, G. Tarjus, M. Tarzia, P. Verrocchio, F. van Wijland.
This work was supported in part by EPSRC Grant No.\ EP/ I017828/1 and Leverhulme Trust Grant No.\ F/00114/BG and by the ERC starting grant NPRGGLASS.


\begin{thebibliography}{99}%

\makeatletter
\providecommand \@ifxundefined [1]{%
 \@ifx{#1\undefined}
}%
\providecommand \@ifnum [1]{%
 \ifnum #1\expandafter \@firstoftwo
 \else \expandafter \@secondoftwo
 \fi
}%
\providecommand \@ifx [1]{%
 \ifx #1\expandafter \@firstoftwo
 \else \expandafter \@secondoftwo
 \fi
}%
\providecommand \natexlab [1]{#1}%
\providecommand \enquote  [1]{``#1''}%
\providecommand \bibnamefont  [1]{#1}%
\providecommand \bibfnamefont [1]{#1}%
\providecommand \citenamefont [1]{#1}%
\providecommand \href@noop [0]{\@secondoftwo}%
\providecommand \href [0]{\begingroup \@sanitize@url \@href}%
\providecommand \@href[1]{\@@startlink{#1}\@@href}%
\providecommand \@@href[1]{\endgroup#1\@@endlink}%
\providecommand \@sanitize@url [0]{\catcode `\\12\catcode `\$12\catcode
  `\&12\catcode `\#12\catcode `\^12\catcode `\_12\catcode `\%12\relax}%
\providecommand \@@startlink[1]{}%
\providecommand \@@endlink[0]{}%
\providecommand \url  [0]{\begingroup\@sanitize@url \@url }%
\providecommand \@url [1]{\endgroup\@href {#1}{\urlprefix }}%
\providecommand \urlprefix  [0]{URL }%
\providecommand \Eprint [0]{\href }%
\providecommand \doibase [0]{http://dx.doi.org/}%
\providecommand \selectlanguage [0]{\@gobble}%
\providecommand \bibinfo  [0]{\@secondoftwo}%
\providecommand \bibfield  [0]{\@secondoftwo}%
\providecommand \translation [1]{[#1]}%
\providecommand \BibitemOpen [0]{}%
\providecommand \bibitemStop [0]{}%
\providecommand \bibitemNoStop [0]{.\EOS\space}%
\providecommand \EOS [0]{\spacefactor3000\relax}%
\providecommand \BibitemShut  [1]{\csname bibitem#1\endcsname}%
\let\auto@bib@innerbib\@empty
%</preamble>

\bibitem[{\citenamefont {Angell}(1995)}]{Angell1995}%
  \BibitemOpen
  \bibfield  {author} {\bibinfo {author} {\bibfnamefont {C.~A.}\ \bibnamefont
  {Angell}},\ }\href@noop {} {\bibfield  {journal} {\bibinfo  {journal}
  {Science}\ }\textbf {\bibinfo {volume} {267}},\ \bibinfo {pages} {1924}
  (\bibinfo {year} {1995})}\BibitemShut {NoStop}%
\bibitem [{\citenamefont {Ediger}\ \emph {et~al.}(1996)\citenamefont {Ediger},
  \citenamefont {Angell},\ and\ \citenamefont {Nagel}}]{Ediger1996}%
  \BibitemOpen
  \bibfield  {author} {\bibinfo {author} {\bibfnamefont {M.}~\bibnamefont
  {Ediger}}, \bibinfo {author} {\bibfnamefont {C.}~\bibnamefont {Angell}}, \
  and\ \bibinfo {author} {\bibfnamefont {S.}~\bibnamefont {Nagel}},\
  }\href@noop {} {\bibfield  {journal} {\bibinfo  {journal} {J. Phys. Chem.}\
  }\textbf {\bibinfo {volume} {100}},\ \bibinfo {pages} {13200} (\bibinfo
  {year} {1996})}\BibitemShut {NoStop}%
\bibitem [{\citenamefont {Angell}\ \emph {et~al.}(2000)\citenamefont {Angell},
  \citenamefont {Ngai}, \citenamefont {McKenna}, \citenamefont {McMillan},\
  and\ \citenamefont {Martin}}]{Angell2000}%
  \BibitemOpen
  \bibfield  {author} {\bibinfo {author} {\bibfnamefont {C.~A.}\ \bibnamefont
  {Angell}}, \bibinfo {author} {\bibfnamefont {K.~L.}\ \bibnamefont {Ngai}},
  \bibinfo {author} {\bibfnamefont {G.~B.}\ \bibnamefont {McKenna}}, \bibinfo
  {author} {\bibfnamefont {P.~F.}\ \bibnamefont {McMillan}}, \ and\ \bibinfo
  {author} {\bibfnamefont {S.~W.}\ \bibnamefont {Martin}},\ }\href@noop {}
  {\bibfield  {journal} {\bibinfo  {journal} {J. Appl. Phys.}\ }\textbf
  {\bibinfo {volume} {88}},\ \bibinfo {pages} {3113} (\bibinfo {year}
  {2000})}\BibitemShut {NoStop}%
\bibitem [{\citenamefont {Debenedetti}\ and\ \citenamefont
  {Stillinger}(2001)}]{Debenedetti2001}%
  \BibitemOpen
  \bibfield  {author} {\bibinfo {author} {\bibfnamefont {P.~G.}\ \bibnamefont
  {Debenedetti}}\ and\ \bibinfo {author} {\bibfnamefont {F.~H.}\ \bibnamefont
  {Stillinger}},\ }\href@noop {} {\bibfield  {journal} {\bibinfo  {journal}
  {Nature}\ }\textbf {\bibinfo {volume} {410}},\ \bibinfo {pages} {259}
  (\bibinfo {year} {2001})}\BibitemShut {NoStop}%
\bibitem [{\citenamefont {Lubchenko}\ and\ \citenamefont
  {Wolynes}(2007)}]{Lubchenko2007}%
  \BibitemOpen
  \bibfield  {author} {\bibinfo {author} {\bibfnamefont {V.}~\bibnamefont
  {Lubchenko}}\ and\ \bibinfo {author} {\bibfnamefont {P.~G.}\ \bibnamefont
  {Wolynes}},\ }\href@noop {} {\bibfield  {journal} {\bibinfo  {journal} {Annu.
  Rev. Phys. Chem.}\ }\textbf {\bibinfo {volume} {58}},\ \bibinfo {pages} {235}
  (\bibinfo {year} {2007})}\BibitemShut {NoStop}%
\bibitem [{\citenamefont {Heuer}(2008)}]{Heuer2008}%
  \BibitemOpen
  \bibfield  {author} {\bibinfo {author} {\bibfnamefont {A.}~\bibnamefont
  {Heuer}},\ }\href@noop {} {\bibfield  {journal} {\bibinfo  {journal} {J.
  Phys: Cond. Matter}\ }\textbf {\bibinfo {volume} {20}},\ \bibinfo {pages}
  {373101} (\bibinfo {year} {2008})}\BibitemShut {NoStop}%
\bibitem [{\citenamefont {Cavagna}(2009)}]{Cavagna2009}%
  \BibitemOpen
  \bibfield  {author} {\bibinfo {author} {\bibfnamefont {A.}~\bibnamefont
  {Cavagna}},\ }\href@noop {} {\bibfield  {journal} {\bibinfo  {journal} {Phys.
  Rep.}\ }\textbf {\bibinfo {volume} {476}},\ \bibinfo {pages} {51} (\bibinfo
  {year} {2009})}\BibitemShut {NoStop}%
\bibitem [{\citenamefont {Chandler}\ and\ \citenamefont
  {Garrahan}(2010)}]{Chandler2010}%
  \BibitemOpen
  \bibfield  {author} {\bibinfo {author} {\bibfnamefont {D.}~\bibnamefont
  {Chandler}}\ and\ \bibinfo {author} {\bibfnamefont {J.~P.}\ \bibnamefont
  {Garrahan}},\ }\href@noop {} {\bibfield  {journal} {\bibinfo  {journal}
  {Annu. Rev. Phys. Chem.}\ }\textbf {\bibinfo {volume} {61}},\ \bibinfo
  {pages} {191} (\bibinfo {year} {2010})}\BibitemShut {NoStop}%
\bibitem [{\citenamefont {Berthier}\ and\ \citenamefont
  {Biroli}(2011)}]{Berthier2011}%
  \BibitemOpen
  \bibfield  {author} {\bibinfo {author} {\bibfnamefont {L.}~\bibnamefont
  {Berthier}}\ and\ \bibinfo {author} {\bibfnamefont {G.}~\bibnamefont
  {Biroli}},\ }\href {\doibase 10.1103/RevModPhys.83.587} {\bibfield  {journal}
  {\bibinfo  {journal} {Rev. Mod. Phys.}\ }\textbf {\bibinfo {volume} {83}},\
  \bibinfo {pages} {587} (\bibinfo {year} {2011})}\BibitemShut {NoStop}%
\bibitem [{\citenamefont {Ediger}(2000)}]{Ediger2000}%
  \BibitemOpen
  \bibfield  {author} {\bibinfo {author} {\bibfnamefont {M.~D.}\ \bibnamefont
  {Ediger}},\ }\href@noop {} {\bibfield  {journal} {\bibinfo  {journal} {Annu.
  Rev. Phys. Chem.}\ }\textbf {\bibinfo {volume} {51}},\ \bibinfo {pages} {99}
  (\bibinfo {year} {2000})}\BibitemShut {NoStop}%
\bibitem [{\citenamefont {Glotzer}(2000)}]{Glotzer2000}%
  \BibitemOpen
  \bibfield  {author} {\bibinfo {author} {\bibfnamefont {S.~C.}\ \bibnamefont
  {Glotzer}},\ }\href@noop {} {\bibfield  {journal} {\bibinfo  {journal} {J.
  Non-Cryst. Solids}\ }\textbf {\bibinfo {volume} {274}},\ \bibinfo {pages}
  {342} (\bibinfo {year} {2000})}\BibitemShut {NoStop}%
\bibitem [{\citenamefont {Andersen}(2005)}]{Andersen2005}%
  \BibitemOpen
  \bibfield  {author} {\bibinfo {author} {\bibfnamefont {H.~C.}\ \bibnamefont
  {Andersen}},\ }\href@noop {} {\bibfield  {journal} {\bibinfo  {journal}
  {Proc. Natl. Acad. Sci. USA}\ }\textbf {\bibinfo {volume} {102}},\ \bibinfo
  {pages} {6686} (\bibinfo {year} {2005})}\BibitemShut {NoStop}%
\bibitem [{\citenamefont {Richert}\ and\ \citenamefont
  {Angell}(1998)}]{Richert1998}%
  \BibitemOpen
  \bibfield  {author} {\bibinfo {author} {\bibfnamefont {R.}~\bibnamefont
  {Richert}}\ and\ \bibinfo {author} {\bibfnamefont {C.}~\bibnamefont
  {Angell}},\ }\href@noop {} {\bibfield  {journal} {\bibinfo  {journal} {J.
  Chem. Phys.}\ }\textbf {\bibinfo {volume} {108}},\ \bibinfo {pages} {9016}
  (\bibinfo {year} {1998})}\BibitemShut {NoStop}%
\bibitem [{\citenamefont {Gottke}\ \emph {et~al.}(2001)\citenamefont {Gottke},
  \citenamefont {David}, \citenamefont {Hinze},\ and\ \citenamefont
  {Fayer}}]{Gottke2001}%
  \BibitemOpen
  \bibfield  {author} {\bibinfo {author} {\bibfnamefont {S.}~\bibnamefont
  {Gottke}}, \bibinfo {author} {\bibfnamefont {D.}~\bibnamefont {David}},
  \bibinfo {author} {\bibfnamefont {G.}~\bibnamefont {Hinze}}, \ and\ \bibinfo
  {author} {\bibfnamefont {M.}~\bibnamefont {Fayer}},\ }\href@noop {}
  {\bibfield  {journal} {\bibinfo  {journal} {J. Phys. Chem. B}\ }\textbf
  {\bibinfo {volume} {105}},\ \bibinfo {pages} {238} (\bibinfo {year}
  {2001})}\BibitemShut {NoStop}%
\bibitem [{\citenamefont {Richert}(2005)}]{Richert2005}%
  \BibitemOpen
  \bibfield  {author} {\bibinfo {author} {\bibfnamefont {R.}~\bibnamefont
  {Richert}},\ }\href@noop {} {\bibfield  {journal} {\bibinfo  {journal} {J.
  Chem. Phys.}\ }\textbf {\bibinfo {volume} {123}},\ \bibinfo {pages} {154502}
  (\bibinfo {year} {2005})}\BibitemShut {NoStop}%
\bibitem [{\citenamefont {Elmatad}\ \emph {et~al.}(2009)\citenamefont
  {Elmatad}, \citenamefont {Chandler},\ and\ \citenamefont
  {Garrahan}}]{Elmatad2009}%
  \BibitemOpen
  \bibfield  {author} {\bibinfo {author} {\bibfnamefont {Y.~S.}\ \bibnamefont
  {Elmatad}}, \bibinfo {author} {\bibfnamefont {D.}~\bibnamefont {Chandler}}, \
  and\ \bibinfo {author} {\bibfnamefont {J.~P.}\ \bibnamefont {Garrahan}},\
  }\href@noop {} {\bibfield  {journal} {\bibinfo  {journal} {J. Phys. Chem. B}\
  }\textbf {\bibinfo {volume} {113}},\ \bibinfo {pages} {5563} (\bibinfo {year}
  {2009})}\BibitemShut {NoStop}%
\bibitem [{\citenamefont {Keys}\ \emph {et~al.}(2011)\citenamefont {Keys},
  \citenamefont {Hedges}, \citenamefont {Garrahan}, \citenamefont {Glotzer},\
  and\ \citenamefont {Chandler}}]{Keys2011}%
  \BibitemOpen
  \bibfield  {author} {\bibinfo {author} {\bibfnamefont {A.~S.}\ \bibnamefont
  {Keys}}, \bibinfo {author} {\bibfnamefont {L.~O.}\ \bibnamefont {Hedges}},
  \bibinfo {author} {\bibfnamefont {J.~P.}\ \bibnamefont {Garrahan}}, \bibinfo
  {author} {\bibfnamefont {S.~C.}\ \bibnamefont {Glotzer}}, \ and\ \bibinfo
  {author} {\bibfnamefont {D.}~\bibnamefont {Chandler}},\ }\href {\doibase
  10.1103/PhysRevX.1.021013} {\bibfield  {journal} {\bibinfo  {journal} {Phys.
  Rev. X}\ }\textbf {\bibinfo {volume} {1}},\ \bibinfo {pages} {021013}
  (\bibinfo {year} {2011})}\BibitemShut {NoStop}%
\bibitem [{\citenamefont {Velikov}\ \emph {et~al.}(2001)\citenamefont
  {Velikov}, \citenamefont {Borick},\ and\ \citenamefont
  {Angell}}]{Velikov2001}%
  \BibitemOpen
  \bibfield  {author} {\bibinfo {author} {\bibfnamefont {V.}~\bibnamefont
  {Velikov}}, \bibinfo {author} {\bibfnamefont {S.}~\bibnamefont {Borick}}, \
  and\ \bibinfo {author} {\bibfnamefont {C.~A.}\ \bibnamefont {Angell}},\
  }\href@noop {} {\bibfield  {journal} {\bibinfo  {journal} {Science}\ }\textbf
  {\bibinfo {volume} {294}},\ \bibinfo {pages} {2335} (\bibinfo {year}
  {2001})}\BibitemShut {NoStop}%
\bibitem [{\citenamefont {Sausset}\ \emph {et~al.}(2010)\citenamefont
  {Sausset}, \citenamefont {Biroli},\ and\ \citenamefont
  {Kurchan}}]{Sausset2010}%
 \BibitemOpen
  \bibfield  {author} {\bibinfo {author} {\bibfnamefont {F.}~\bibnamefont
  {Sausset}}, \bibinfo {author} {\bibfnamefont {G.}~\bibnamefont {Biroli}}, \
  and\ \bibinfo {author} {\bibfnamefont {J.}~\bibnamefont {Kurchan}},\
  }\href@noop {} {\bibfield  {journal} {\bibinfo  {journal} {J. Stat. Phys.}\
  }\textbf {\bibinfo {volume} {140}},\ \bibinfo {pages} {718} (\bibinfo {year}
  {2010})}\BibitemShut {NoStop}%
\bibitem{KTW}
T. R. Kirkpatrick, D. Thirumalai, and P. G. Wolynes, Phys. Rev. A {\bf 40} (1989) 1045.

\bibitem [{\citenamefont {Gotze}\ and\ \citenamefont
  {Sjorgen}(1992)}]{Gotze1992}%
  \BibitemOpen
  \bibfield  {author} {\bibinfo {author} {\bibfnamefont {W.}~\bibnamefont
  {Gotze}}\ and\ \bibinfo {author} {\bibfnamefont {L.}~\bibnamefont
  {Sjorgen}},\ }\href@noop {} {\bibfield  {journal} {\bibinfo  {journal} {Rep.
  Prog. Phys.}\ }\textbf {\bibinfo {volume} {55}},\ \bibinfo {pages} {241}
  (\bibinfo {year} {1992})}\BibitemShut {NoStop}%
\bibitem [{\citenamefont {Reichman}\ and\ \citenamefont
  {Charbonneau}(2005)}]{Reichman2005}%
  \BibitemOpen
  \bibfield  {author} {\bibinfo {author} {\bibfnamefont {D.}~\bibnamefont
  {Reichman}}\ and\ \bibinfo {author} {\bibfnamefont {P.}~\bibnamefont
  {Charbonneau}},\ }\href@noop {} {\bibfield  {journal} {\bibinfo  {journal}
  {J. Stat. Mech.}\ }\textbf {\bibinfo {volume} {2005}},\ \bibinfo {pages}
  {P05013} (\bibinfo {year} {2005})}\BibitemShut {NoStop}%
\bibitem [{\citenamefont {Weeks}\ \emph {et~al.}(2000)\citenamefont {Weeks},
  \citenamefont {Crocker}, \citenamefont {Levitt}, \citenamefont {Schofield},\
  and\ \citenamefont {Weitz}}]{Weeks2000}%
  \BibitemOpen
  \bibfield  {author} {\bibinfo {author} {\bibfnamefont {E.~R.}\ \bibnamefont
  {Weeks}}, \bibinfo {author} {\bibfnamefont {J.~C.}\ \bibnamefont {Crocker}},
  \bibinfo {author} {\bibfnamefont {A.~C.}\ \bibnamefont {Levitt}}, \bibinfo
  {author} {\bibfnamefont {A.}~\bibnamefont {Schofield}}, \ and\ \bibinfo
  {author} {\bibfnamefont {D.~A.}\ \bibnamefont {Weitz}},\ }\href@noop {}
  {\bibfield  {journal} {\bibinfo  {journal} {Science}\ }\textbf {\bibinfo
  {volume} {287}},\ \bibinfo {pages} {627} (\bibinfo {year}
  {2000})}\BibitemShut {NoStop}%
\bibitem [{\citenamefont {Kegel}\ and\ \citenamefont {van
  Blaaderen}(2000)}]{Kegel2000}%
  \BibitemOpen
  \bibfield  {author} {\bibinfo {author} {\bibfnamefont {W.~K.}\ \bibnamefont
  {Kegel}}\ and\ \bibinfo {author} {\bibfnamefont {A.}~\bibnamefont {van
  Blaaderen}},\ }\href@noop {} {\bibfield  {journal} {\bibinfo  {journal}
  {Science}\ }\textbf {\bibinfo {volume} {287}},\ \bibinfo {pages} {290}
  (\bibinfo {year} {2000})}\BibitemShut {NoStop}%
\bibitem [{\citenamefont {Dauchot}\ \emph {et~al.}(2005)\citenamefont
  {Dauchot}, \citenamefont {Marty},\ and\ \citenamefont
  {Biroli}}]{Dauchot2005}%
  \BibitemOpen
  \bibfield  {author} {\bibinfo {author} {\bibfnamefont {O.}~\bibnamefont
  {Dauchot}}, \bibinfo {author} {\bibfnamefont {G.}~\bibnamefont {Marty}}, \
  and\ \bibinfo {author} {\bibfnamefont {G.}~\bibnamefont {Biroli}},\ }\href
  {\doibase 10.1103/PhysRevLett.95.265701} {\bibfield  {journal} {\bibinfo
  {journal} {Phys. Rev. Lett.}\ }\textbf {\bibinfo {volume} {95}},\ \bibinfo
  {eid} {265701} (\bibinfo {year} {2005})}\BibitemShut {NoStop}%
\bibitem [{\citenamefont {Keys}\ \emph {et~al.}(2007)\citenamefont {Keys},
  \citenamefont {Abate}, \citenamefont {Glotzer},\ and\ \citenamefont
  {Durian}}]{Keys2007}%
  \BibitemOpen
  \bibfield  {author} {\bibinfo {author} {\bibfnamefont {A.}~\bibnamefont
  {Keys}}, \bibinfo {author} {\bibfnamefont {A.}~\bibnamefont {Abate}},
  \bibinfo {author} {\bibfnamefont {S.}~\bibnamefont {Glotzer}}, \ and\
  \bibinfo {author} {\bibfnamefont {D.}~\bibnamefont {Durian}},\ }\href@noop {}
  {\bibfield  {journal} {\bibinfo  {journal} {Nature Phys.}\ }\textbf {\bibinfo
  {volume} {3}},\ \bibinfo {pages} {260} (\bibinfo {year} {2007})}\BibitemShut
  {NoStop}%
\bibitem [{\citenamefont {Mayer}\ \emph {et~al.}(2004)\citenamefont {Mayer},
  \citenamefont {Bissig}, \citenamefont {Berthier}, \citenamefont {Cipelletti},
  \citenamefont {Garrahan}, \citenamefont {Sollich},\ and\ \citenamefont
  {Trappe}}]{Mayer2004}%
  \BibitemOpen
  \bibfield  {author} {\bibinfo {author} {\bibfnamefont {P.}~\bibnamefont
  {Mayer}}, \bibinfo {author} {\bibfnamefont {H.}~\bibnamefont {Bissig}},
  \bibinfo {author} {\bibfnamefont {L.}~\bibnamefont {Berthier}}, \bibinfo
  {author} {\bibfnamefont {L.}~\bibnamefont {Cipelletti}}, \bibinfo {author}
  {\bibfnamefont {J.}~\bibnamefont {Garrahan}}, \bibinfo {author}
  {\bibfnamefont {P.}~\bibnamefont {Sollich}}, \ and\ \bibinfo {author}
  {\bibfnamefont {V.}~\bibnamefont {Trappe}},\ }\href@noop {} {\bibfield
  {journal} {\bibinfo  {journal} {Phys. Rev. Lett.}\ }\textbf {\bibinfo
  {volume} {93}},\ \bibinfo {pages} {115701} (\bibinfo {year}
  {2004})}\BibitemShut {NoStop}%
\bibitem [{\citenamefont {Brun}\ \emph {et~al.}(2013)\citenamefont {Brun},
  \citenamefont {Ladieu}, \citenamefont {Biroli}, \citenamefont {Bouchaud}
  \emph {et~al.}}]{Brun2013}%
  \BibitemOpen
  \bibfield  {author} {\bibinfo {author} {\bibfnamefont {C.}~\bibnamefont
  {Brun}}, \bibinfo {author} {\bibfnamefont {F.}~\bibnamefont {Ladieu}},
  \bibinfo {author} {\bibfnamefont {G.}~\bibnamefont {Biroli}}, \bibinfo
  {author} {\bibfnamefont {J.}~\bibnamefont {Bouchaud}},  \emph {et~al.},\
  }\href@noop {} {\bibfield  {journal} {\bibinfo  {journal} {in press Phys. Rev. Lett.}\
  } (\bibinfo {year} {2013})}\BibitemShut {NoStop}%
\bibitem [{\citenamefont {Angelini}\ \emph {et~al.}(2011)\citenamefont
  {Angelini}, \citenamefont {Hannezo}, \citenamefont {Trepat}, \citenamefont
  {Marquez}, \citenamefont {Fredberg},\ and\ \citenamefont
  {Weitz}}]{Angelini2011}%
  \BibitemOpen
  \bibfield  {author} {\bibinfo {author} {\bibfnamefont {T.}~\bibnamefont
  {Angelini}}, \bibinfo {author} {\bibfnamefont {E.}~\bibnamefont {Hannezo}},
  \bibinfo {author} {\bibfnamefont {X.}~\bibnamefont {Trepat}}, \bibinfo
  {author} {\bibfnamefont {M.}~\bibnamefont {Marquez}}, \bibinfo {author}
  {\bibfnamefont {J.}~\bibnamefont {Fredberg}}, \ and\ \bibinfo {author}
  {\bibfnamefont {D.}~\bibnamefont {Weitz}},\ }\href@noop {} {\bibfield
  {journal} {\bibinfo  {journal} {Proc. Natl. Acad. Sci. USA}\ }\textbf
  {\bibinfo {volume} {108}},\ \bibinfo {pages} {4714} (\bibinfo {year}
  {2011})}\BibitemShut {NoStop}%
\bibitem [{\citenamefont {Schweizer}(2007)}]{Schweizer2007}%
  \BibitemOpen
  \bibfield  {author} {\bibinfo {author} {\bibfnamefont {K.}~\bibnamefont
  {Schweizer}},\ }\href@noop {} {\bibfield  {journal} {\bibinfo  {journal}
  {Curr. Opin. Colloid Interface Sci.}\ }\textbf {\bibinfo {volume} {12}},\
  \bibinfo {pages} {297} (\bibinfo {year} {2007})}\BibitemShut {NoStop}%
\bibitem [{\citenamefont {Trachenko}\ and\ \citenamefont
  {Brazhkin}(2011)}]{Trachenko2011}%
  \BibitemOpen
  \bibfield  {author} {\bibinfo {author} {\bibfnamefont {K.}~\bibnamefont
  {Trachenko}}\ and\ \bibinfo {author} {\bibfnamefont {V.~V.}\ \bibnamefont
  {Brazhkin}},\ }\href@noop {} {\bibfield  {journal} {\bibinfo  {journal}
  {Phys. Rev. B}\ }\textbf {\bibinfo {volume} {83}} (\bibinfo {year}
  {2011})}\BibitemShut {NoStop}%
  \bibitem{wyartcp}
  M. Wyart, Phys. Rev. Lett. {\bf 104} (2010) 095901.
\bibitem [{\citenamefont {Van~Hecke}(2009)}]{Van-Hecke2009}%
  \BibitemOpen
  \bibfield  {author} {\bibinfo {author} {\bibfnamefont {M.}~\bibnamefont
  {Van~Hecke}},\ }\href@noop {} {\bibfield  {journal} {\bibinfo  {journal} {J.
  Phys: Cond. Matter}\ }\textbf {\bibinfo {volume} {22}},\ \bibinfo {pages}
  {033101} (\bibinfo {year} {2009})}\BibitemShut {NoStop}%
\bibitem [{\citenamefont {Liu}\ \emph {et~al.}(2010)\citenamefont {Liu},
  \citenamefont {Nagel}, \citenamefont {Van~Saarloos},\ and\ \citenamefont
  {Wyart}}]{Liu2010}%
  \BibitemOpen
  \bibfield  {author} {\bibinfo {author} {\bibfnamefont {A.}~\bibnamefont
  {Liu}}, \bibinfo {author} {\bibfnamefont {S.}~\bibnamefont {Nagel}}, \bibinfo
  {author} {\bibfnamefont {W.}~\bibnamefont {Van~Saarloos}}, \ and\ \bibinfo
  {author} {\bibfnamefont {M.}~\bibnamefont {Wyart}},\ }\href@noop {}
  {\bibfield  {journal} {\bibinfo  {journal} {arXiv:1006.2365}\ } (\bibinfo
  {year} {2010})}\BibitemShut {NoStop}%
\bibitem{procaccia}
S. Karmakar, E. Lerner, I. Procaccia, 
Phys. Rev. E {\bf 82} 055103(R) (2010), Phys. Rev. E {\bf 82} 026105 (2010).
H.G.E. Hentschel, S. Karmakar, E. Lerner, I. Procaccia, Phys. Rev. E {\bf 83} 061101 (2011).
\bibitem [{\citenamefont {Torquato}\ and\ \citenamefont
  {Stillinger}(2010)}]{Torquato2010}%
  \BibitemOpen
  \bibfield  {author} {\bibinfo {author} {\bibfnamefont {S.}~\bibnamefont
  {Torquato}}\ and\ \bibinfo {author} {\bibfnamefont {F.}~\bibnamefont
  {Stillinger}},\ }\href@noop {} {\bibfield  {journal} {\bibinfo  {journal}
  {Rev. Mod. Phys.}\ }\textbf {\bibinfo {volume} {82}},\ \bibinfo {pages}
  {2633} (\bibinfo {year} {2010})}\BibitemShut {NoStop}  
\bibitem{barrat}
J.-L. Barrat and A. Lemaitre, {\em Heterogeneities in amorphous systems under shear}, in \cite{bookDH}.
\bibitem{bookDH}
{\it Dynamical heterogeneities in glasses, colloids and granular materials}, Eds. L. Berthier, G. Biroli, J.-P. Bouchaud, L. Cipelletti, W. van Saarloos, Oxford University Press (2011).

\bibitem{garrahanquantum}
B. Olmos, I. Lesanovsky, and J. P. Garrahan, Phys. Rev. Lett. 
{\bf 109} 020403 (2012).

\bibitem{nussinov}
Z. Nussinov, P. Johnson, M. J. Graf, A. V. Balatsky, 
{\it Mapping between finite temperature classical and zero temperature quantum systems: quantum critical jamming and quantum dynamical heterogeneities}, arXiv:1209.3823.

\bibitem{Hodgdon1993}
J. A. Hodgdon and F. H. Stillinger, Phys. Rev. E {\bf 48},
207 (1993).

\bibitem{Tarjus1995}
G. Tarjus and D. Kivelson, J. Chem. Phys. {\bf 103}, 3071
(1995).

\bibitem{Tarjus2011}
G. Tarjus, {\em An overview of the theories of the glass transition}, in \cite{bookDH}.

\bibitem{BBBJ2011}
L. Berthier, G. Biroli, J.-P. Bouchaud, R.L. Jack, ÒOverview of different characterisations of dynamic heterogeneity,Ó in \cite{bookDH}.

\bibitem{Harrowell2011}
P. Harrowell, {\em The Length Scales of Dynamic Heterogeneity: Results from Molecular Dynamics Simulations}, in \cite{bookDH}.

\bibitem{Richert2011}
R. Richert, N. Israeloff, C. Alba-Simionesco, F. Ladieu and D. L'H\^ote, {\em Experimental Approaches to Heterogeneous Dynamics}, in \cite{bookDH}.

\bibitem{CDB}
R. Candelier, O. Dauchot, G. Biroli
Phys. Rev. Lett. {\bf 102} (2009) 088001.

\bibitem{harrowell}
P. Harrowell, Phys.Rev.E {\bf 48} (1993) 4359; 
Perera and P. Harrowell, Phys. Rev. E {\bf 54} (1996) 1652. 

\bibitem{Garrahan2002}
J.P. Garrahan and D. Chandler,
Phys. Rev. Lett. {\bf 89}, 035704 (2002).


\bibitem[{\citenamefont{Ritort and Sollich}(2003)}]{Ritort2003}
\bibinfo{author}{\bibfnamefont{F.}~\bibnamefont{Ritort}} \bibnamefont{and}
  \bibinfo{author}{\bibfnamefont{P.}~\bibnamefont{Sollich}},
  \bibinfo{journal}{Adv. Phys.} \textbf{\bibinfo{volume}{52}},
  \bibinfo{pages}{219} (\bibinfo{year}{2003}).

\bibitem{wolynesdf}
S. M. Bhattacharyya, B. Bagchi, and P. G. Wolynes, 
PNAS {\bf 105} (2008) 16077.

\bibitem{harrowelldauchot}
R. Candelier, A. Widmer-Cooper et al.,
 Phys. Rev. Lett. {\bf 105} (2010) 135702.
 
 \bibitem{propensity}
A. Widmer-Cooper,
P. Harrowell, and H. Fynewever, Phys. Rev. Lett. {\bf 93} (2004) 135701.

\bibitem{harrowellcooper}
A. Widmer-Cooper, P. Harrowell Phys. Rev. Lett. {\bf 96} (2006) 185701;
J. Phys.: Condens. Matter {\bf 17} (2005) S4025.

\bibitem{harrowellreichman}
A. Widmer-Cooper, H. Perry, P. Harrowell, D. R. Reichman,
Nature Physics {\bf 4} (2008) 711.

\bibitem{Elmatad2012}
Y.S. Elmatad and A.S. Keys,
Phys. Rev. E {\bf 85}, 061502 (2012).

\bibitem{wyart}
C. Brito and M. Wyart, J. Chem. Phys. {\bf 131} 024504 (2009).


\bibitem{Ashton2009}
D. Ashton and J. P. Garrahan
Euro. Phys. J. EÊ {\bf 30},Ê 303Ê (2009).
 
\bibitem{Charbonneautarjus}
B. Charbonneau, P. Charbonneau, G. Tarjus, Journal of Chem. Phys. {\bf 138}, 12A515 (2013).

\bibitem{royal}
T. Speck, A. Malins, and C. P. Royall, Phys. Rev. Lett. {\bf 109}, 195703.

\bibitem{tanaka}
A. Malins et al., {\it arXiv:1203.1732}. 

\bibitem{levinekurchan}
Kurchan J. and Levine D., J. Phys. A, {\bf 44} (2011) 035001.

\bibitem{cammarotaPR}
C. Cammarota, G. Biroli, EuroPhys. Lett. Vol. {\bf 98} 36005 (2012).

\bibitem{Bouchaud2004}
J.-P. Bouchaud, G. Biroli, J. Chem. Phys. {\bf 121}, 7347 (2004).

\bibitem{CGV}
A. Cavagna, T. S. Grigera, and P. Verrocchio, Phys. Rev. Lett. {\bf 98}, 187801 (2007).

\bibitem{BBCGV}
G. Biroli, J.-P. Bouchaud, A. Cavagna, T. S. Grigera, and P. Verrocchio, Nature Physics 4, 771 (2008).

\bibitem{refsgeometries1}
F. Sausset and G. Tarjus, Phys. Rev. Lett. {\bf 104}, 065701 (2010).
\bibitem{refsgeometries2}
B. Charbonneau, P. Charbonneau, and G. Tarjus, Phys. Rev. Lett. {\bf 108}, 035701 (2012).
\bibitem{refsgeometries3}
L. Berthier and W. Kob, Phys. Rev. E {\bf 85}, 011102 (2012).
\bibitem{refsgeometries4}
G. M. Hocky, T. E. Markland, and D. R. Reichman, Phys. Rev Lett. {\bf 108}, 225506 (2012).

\bibitem{montanarisemerjian}
A. Montanari, G. Semerjian, J. Stat. Phys. {\bf 125} 23 (2006).

\bibitem{dynstat}
W. Kob, S. Roldan-Vargas, L. Berthier,
Nature Physics {\bf 8}, 697 (2012).

\bibitem{dynstat2} 
P. Charbonneau, G. Tarjus, {\it arXiv:1211.4821}.

\bibitem{Garrahan2011}
J.P. Garrahan, P. Sollich and C. Toninelli,
{\em Kinetically Constrained Models}, in \cite{bookDH}.

\bibitem{Franz2011}
S. Franz and G. Semerjian, {\em Analytical approaches to time and length scales in models of glasses}, in \cite{bookDH}.

\bibitem{franzkac}
S. Franz J. Stat. Mech. (2005) P04001.

\bibitem{wolynesschmalian}
M. Dzero, J. Schmalian, and P. G. Wolynes, Phys. Rev. B, Rapid Comm. {\bf 72} (2005) 100201.

\bibitem{jackgarrahan}
R.L. Jack, J.P. Garrahan, J Chem Phys. {\bf 123} (2005)164508.

\bibitem{pusey} 
P. N. Pusey and  W. van Megen, Nature {\bf 320}, 340 (1986).

\bibitem{dauchot}
R. Candelier, O. Dauchot, G. Biroli, EuroPhys. Lett. {\bf 92} 24003 (2010).

\bibitem{bernal}
J.D. Bernal, Nature {\bf 185} 68 (1960).

\bibitem{torquato}
S. Torquato, T.M. Truskett, P.G. Debenedetti, 
Phys. Rev. Lett. {\bf 84} (2000) 2064.

\bibitem{liunagel}
A. J. Liu and S. R. Nagel, Nature {\bf 396}, 21 (1998).

\bibitem{berthierpointjs}
P. Chaudhuri, L. Berthier, S. Sastry, Phys. Rev. Lett. {\bf 104}, 165701 (2010).

\bibitem{krzakalakurchan}
R. Mari, F. Krzakala, J. Kurchan, Phys. Rev. Lett. {\bf 103} 025701 (2009).

\bibitem{berthierwitten}
L. Berthier, T.A. Witten, Phys. Rev. E {\bf 80} 021502 (2009); 
Europhys. Lett. {86} 10001 (2009).

\bibitem{berthiersollich}
A. Ikeda, L. Berthier, and P. Sollich,  Phys. Rev. Lett. {\bf 109} 018301 (2012).

\bibitem{RFOTLW}
V. Lubchenko and P.G. Wolynes in ``Structural Glasses and Supercooled Liquids: Theory, Experiment, and Applications'', Wiley 2012 (Eds P.G. Wolynes and V. Lubchenko).

\bibitem{RFOTBB}
G. Biroli and J.-P. Bouchaud in ``Structural Glasses and Supercooled Liquids: Theory, Experiment, and Applications'', Wiley 2012 (Eds P.G. Wolynes and V. Lubchenko). 


\bibitem{IMCT}
G. Biroli, J.P. Bouchaud, K. Miyazaki, D.R. Reichman
Phys. Rev. Lett. {\bf 97} (2006) 195701.
 
\bibitem{JCPlong}
 L. Berthier, G. Biroli, J.-P. Bouchaud, W. Kob, K. Miyazaki, and D. R. Reichman, 
 J. Chem. Phys. {\bf 126} (2007) 184503; J. Chem. Phys. {\bf 126} (2007) 184504.
 
\bibitem{fluctpapersparisi} 
 S. Franz et al.,  J. Chem. Phys. {\bf 138} (2013) 12A540.
 S. Franz, G. Parisi, F. Ricci-Tersenghi, T. Rizzo, Eur. Phys. J. E. {\bf 34} (2011) 102.

\bibitem{dysonrem}
M. Castellana, A. Decelle, S. Franz, M. M\'ezard and G. Parisi. Phys. Rev. Lett. {\bf 104}, (2010) 127206. 

\bibitem{MKFT}
C. Cammarota, G. Biroli, M. Tarzia, and G. Tarjus,
Phys. Rev. Lett. {\bf 106} (2011) 115705.
  
  \bibitem{cavagna}
  C. Cammarota, A. Cavagna, G. Gradenigo, T. S. Grigera, and P. Verrocchio, J. Chem. Phys. {\bf 131} 194901 (2009).
  
  \bibitem{sastry}
  S. Karmakar, C. Dasgupta and S. Sastry, Proc. Natl. Acad. Sci. (US) {\bf 106} (2009) 3675.
  
  \bibitem{parisizamponi}
  G. Parisi and F. Zamponi, Rev. Mod. Phys. {\bf 82} (2010) 789.
  
  
\bibitem{Weeks1971} J.D. Weeks, D. Chandler, H.C. Andersen, J. Chem. Phys. {\bf 54}, 5237 (1971).

  
  \bibitem{ikedaberthierbiroli}
  A. Ikeda, L. Berthier, and G. Biroli, J. Chem. Phys. {\bf 138} (2013) 12A507.
  
   \bibitem{kurchanparisizamponi}
  J.Kurchan, G.Parisi, F.Zamponi  J.Stat.Mech. (2012) P10012. 
  
  \bibitem{wolyneswalcwack}
  J. D. Stevenson, A. M. Walczak, R. W. Hall, and P. G. Wolynes, J. Chem. Phys. {\bf 129} (2008) 194505. 
  
  \bibitem{fuchscates}
  J.M. Brader, Th. Voigtmann, M. Fuchs, R.G. Larson, M.E. Cates, 
  PNAS {\bf 106} 15186 (2009).
  
  \bibitem{reichmanmiyazaki}
  K. Miyazaki, D.R. Reichman, R. Yamamoto, 
  Phys. Rev. E {\bf 70} (2004) 011501.
  
\bibitem{Jung2004}
Y.~Jung, J.~P.~Garrahan and D.~Chandler, Phys. Rev. E {\bf 69}, 061205 (2004); J. Chem. Phys. {\bf 123}, 
084509 (2005)

\bibitem{Berthier2005}
L. Berthier, D. Chandler and J.P. Garrahan, Europhys. Lett. {\bf 69}, 
320 (2005); D. Chandler et al., Phys. Rev. E {\bf74}, 051501 (2006). 

\bibitem{Barkai2004}
E. Barkai and Y. J. Jung and R. Silbey, 
Annu. Rev. Phys. Chem.Ê {\bf 55},Ê 457Ê (2004).

\bibitem{Montroll1965}
E.W. Montroll and G.H. Weiss, J. Math. Phys. {\bf 6}, 167 (1965). 

\bibitem{Hedges2007}
L.O. Hedges, L. Maibaum, D. Chandler and J.P. Garrahan, J. Chem. Phys. {\bf 127}, 211101 (2007).

\bibitem{Chaudhuri2007}
P. Chaudhuri, L. Berthier and W. Kob, Phys. Rev. Lett. {\bf99}, 060604 (2007).  
  
  \bibitem{BBSE}
  G. Biroli, J.-P. Bouchaud, J. Phys. Cond. Matt. {\bf 19}, 205101 (2007).
  
  \bibitem{XiaWolynes}
X. Xia, P. G. Wolynes, J. Phys. Chem. B {\bf 105}, 6570 (2001).

\bibitem{charbonneau}
P. Charbonneau, G. Parisi and F. Zamponi, arXiv:1210.6073.

\bibitem{MCThighD1}
A. Ikeda, K. Miyazaki, Phys. Rev. Lett. {\bf 104} (2010) 255704.
\bibitem{MCThighD2}
B. Schmid, R. Schilling, Phys. Rev. E {\bf 81} (2010) 041502.

\bibitem{Tarjus2005}
G. Tarjus, S.A. Kivelson, Z. Nussinov, P. Viot, J. Phys. Cond. Mat. {\bf 17} R1143 (2005).

\bibitem{LD}
For reviews see, J.-P. Eckmann and D. Ruelle,
   Rev. Mod. Phys. {\bf 57}, 617 (1985);
   H. Touchette, Phys. Rep. {\bf 478}, 1 (2009).

\bibitem{Lecomte2005}
V. Lecomte, C. Appert-Rolland and F.  van Wijland, Phys. Rev. Lett. {\bf 95}, 010601 (2005);  J. Stat. Phys. {\bf 127}, 51 (2007).

\bibitem{Merolle2005}
M. Merolle and J. Garrahan and D. Chandler,
Proc. Natl. Acad. Sci. USAÊ {\bf 102},Ê 10837Ê (2005).

\bibitem{Garrahan2007}
J.P. Garrahan, R.L. Jack, V. Lecomte, E. Pitard, K. van Duijvendijk, F. van Wijland,
Phys. Rev. Lett. {\bf 98}, 195702 (2007).

\bibitem{Hedges2009}
L.O. Hedges, R.L. Jack, J.P. Garrahan and D. Chandler, Science {\bf 323}, 1309 (2009).

\bibitem{Speck2012}.
T. Speck and D. Chandler, J. Chem. Phys. {\bf 136}, 184509 (2012).

\bibitem{Fullerton2013}
C.J. Fullerton and R.L. Jack, arXiv:1302.6880.

\bibitem{Elmatad2010}
Y. S. Elmatad and R. L. Jack and D. Chandler and J. P. Garrahan,
Proc. Natl. Acad. Sci. USA {\bf 107},Ê 12793Ê (2010).

\bibitem{PNAS}
C. Cammarota and G. Biroli, Proc. Natl. Acad. Sci. 109,
(2012) 8850.

\bibitem{PNASlong}
C. Cammarota, G. Biroli, J. Chem. Phys. special issue on glass transition, arXiv 1210.8399.

\bibitem{FPRFIM}
S. Franz, G. Parisi and F. Ricci-Tersenghi, JSTAT (2013) L02001.

\bibitem{cammarotaepl}
C. Cammarota, arXiv:1211.4001, to appear on EuroPhys. Lett. 

\bibitem{kobberthierpinning}
W. Kob and L. Berthier, {\it Probing a liquid to glass transition in equilibrium}, 
arXiv: 1301.1795. 

\bibitem{jackberthier}
R.L. Jack and L. Berthier, Phys. Rev. E {\bf 85} (2012) 021120.

\bibitem{kim1}
K. Kim, Europhys. Lett. {\bf 61} (2003) 790.

\bibitem{kim2}
K. Kim, K. Miyazaki, and S. Saito, J. Phys.: Condens. Matter {\bf 23} (2011) 234123.

\bibitem{karmakarprocaccia}
S. Karmakar, I. Procaccia, 
{\it  Exposing the static scale of the glass transition by random pinning}, 
arXiv:1105.4053

\bibitem{FP}
S. Franz and G. Parisi Phys. Rev. Lett. {\bf 79} 2486 (1997); Physica A {\bf 261} 317 (1998).

\bibitem{cavagnacammarota}
C. Cammarota, A. Cavagna, I. Giardina, G. Gradenigo, T. S. Grigera, G. Parisi, and
P. Verrocchio, Phys. Rev. Lett. {\bf 105} 055703 (2010).

\bibitem{Flindt2013}
C. Flindt and J.P. Garrahan,
Phys. Rev. Lett. {\bf 110}, 050601 (2013).

\bibitem{Swallen2007}
S.F. Swallen et al.\, Science {\bf 315}, 353 (2007).

\bibitem{Swallen2009}
S.F. Swallen, K. Traynor, R.J. McMahon, and M. D. Ediger 
Phys. Rev. Lett. {\bf 102}  065503 (2009).

\bibitem{Sepulveda2013}
A. Sepulveda, S.F. Swallen and M.D. Ediger, J. Chem. Phys. {\bf 138}, 12 (2013).

\bibitem{Cugliandolo2002}
L.F. Cugliandolo, 
Lecture notes in Slow Relaxation and non equilibrium dynamics in condensed matter, Les Houches Session 77 July 2002, Ê
J-L Barrat, J Dalibard, J Kurchan, M V Feigel'man eds.Ê

\bibitem{Corberi2011}
F. Corberi, L.F. Cugliandolo and H. Yoshino, {\em Growing length scales in aging systems}, in \cite{bookDH}.

\bibitem{wolynesaging}
V. Lubchenko and P.G. Wolynes, 
J. Chem. Phys. {\bf 121}, 2852 (2004).

\bibitem{Keys2013}
A.S. Keys, J.P. Garrahan and D. Chandler, in press Proc. Natl. Acad. Sci. USA (2013).

\bibitem{Haroche2006}
S. Haroche and J.-M. Raimond, Exploring the Quantum, (Oxford University Press, 2006). 

\bibitem{Bloch2008}
I. Bloch, J. Dalibard and W. Zwerger, Rev. Mod. Phys. {\bf 80}, 885 (2008). 

\bibitem{Liebfried2003} D. Leibfried, R. Blatt, C. Monroe and D. Wineland, Rev. Mod. Phys. {\bf 75}, 281 (2003). 

\bibitem{Schoelkopf2008} R.J. Schoelkopf and S. M. Girvin, Nature {\bf 451}, 664 (2008).

\bibitem{LaHaye2009} M.D. LaHaye et al., Nature {\bf 459}, 960 (2009).

\bibitem{Prokofiev}
M. Boninsegni, N. Prokof'ev, and B. Svistunov, Phys. Rev. Lett. {\bf 96} (2006) 105301.
\bibitem[{\citenamefont{Biroli et~al.}(2008)\citenamefont{Biroli, Chamon, and
  Zamponi}}]{Biroli08}
\bibinfo{author}{\bibfnamefont{G.}~\bibnamefont{Biroli}},
  \bibinfo{author}{\bibfnamefont{C.}~\bibnamefont{Chamon}}, \bibnamefont{and}
  \bibinfo{author}{\bibfnamefont{F.}~\bibnamefont{Zamponi}},
  \bibinfo{journal}{Phys. Rev. B} \textbf{\bibinfo{volume}{78}},
  \bibinfo{pages}{224306} (\bibinfo{year}{2008}).

\bibitem{davis}
 A.V. Balatsky, M. Yamashita, S. Davis, Science {\bf 324} (2009) 632.

\bibitem[{\citenamefont{Das and Chakrabarti}(2008)}]{Das08}
\bibinfo{author}{\bibfnamefont{A.}~\bibnamefont{Das}} \bibnamefont{and}
  \bibinfo{author}{\bibfnamefont{B.~K.} \bibnamefont{Chakrabarti}},
  \bibinfo{journal}{Rev. Mod. Phys.} \textbf{\bibinfo{volume}{80}},
  \bibinfo{pages}{1061} (\bibinfo{year}{2008}).
  
  \bibitem{reviewMF}
  V. Bapst, L. Foini, F. Krzakala, G. Semerjian, F. Zamponi,
   Physics Reports {\bf 523} (2013) 127. 

\bibitem[{\citenamefont{Amir et~al.}(2009)\citenamefont{Amir, Oreg, and
  Imry}}]{Amir09}
\bibinfo{author}{\bibfnamefont{A.}~\bibnamefont{Amir}},
  \bibinfo{author}{\bibfnamefont{Y.}~\bibnamefont{Oreg}}, \bibnamefont{and}
  \bibinfo{author}{\bibfnamefont{Y.}~\bibnamefont{Imry}},
  \bibinfo{journal}{Phys. Rev. Lett.} \textbf{\bibinfo{volume}{103}},
  \bibinfo{pages}{126403} (\bibinfo{year}{2009}).

\bibitem{Polkovnikov2011}
A. Polkovnikov, K. Sengupta, A. Silva and M. Vengalattore, Rev. Mod. Phys. {\bf 83}, 863 (2011).


\bibitem{Altshuler}
D.M. Basko, I. Aleiner, B.L. Altshuler, Ann. Phys. {\bf 321} (2006) 1126.

\bibitem{WolynesMB} D.E. Logan and P.G. Wolynes, J. Chem. Phys. {\bf 93}, 4994 (1990);
R. Bigwood {\it et al.}
Proc. Nat. Acad. Sci. {\bf 95}, 5960 (1998).


\bibitem[{\citenamefont{Pal and Huse}(2010)}]{Pal2010}
\bibinfo{author}{\bibfnamefont{A.}~\bibnamefont{Pal}} \bibnamefont{and}
  \bibinfo{author}{\bibfnamefont{D.~A.} \bibnamefont{Huse}},
  \bibinfo{journal}{Phys. Rev. B} \textbf{\bibinfo{volume}{82}},
  \bibinfo{pages}{174411} (\bibinfo{year}{2010}).


\bibitem[{\citenamefont{Markland et~al.}(2011)\citenamefont{Markland, Morrone,
  Berne, Miyazaki, Rabani, and Reichman}}]{Markland2011}
\bibinfo{author}{\bibfnamefont{T.~E.} \bibnamefont{Markland}},
  \bibinfo{author}{\bibfnamefont{J.~A.} \bibnamefont{Morrone}},
  \bibinfo{author}{\bibfnamefont{B.~J.} \bibnamefont{Berne}},
  \bibinfo{author}{\bibfnamefont{K.}~\bibnamefont{Miyazaki}},
  \bibinfo{author}{\bibfnamefont{E.}~\bibnamefont{Rabani}}, \bibnamefont{and}
  \bibinfo{author}{\bibfnamefont{D.~R.} \bibnamefont{Reichman}},
  \bibinfo{journal}{Nat. Phys.} \textbf{\bibinfo{volume}{7}},
  \bibinfo{pages}{134} (\bibinfo{year}{2011}).

\bibitem{Chamon2005} C. Chamon, Phys. Rev. Lett. {\bf 94}, 040402 (2005)

\bibitem{Foini2011} L. Foini, G. Semerjian and F. Zamponi, Phys. Rev. B {\bf 83}, 094513 (2011)

\bibitem{Nussinov2008} Z. Nussinov, Physics {\bf 1}, 40 (2008)

\bibitem{Poletti2012}
D. Poletti, P. Barmettler, A. Georges and C. Kollath, 
arXiv:1212.4637.



\end{thebibliography}
\end{document}